\documentclass{aa}  
\usepackage{graphicx}
\usepackage{txfonts}
\usepackage[varvw]{newtxmath}	

\usepackage{geometry}
\usepackage{xcolor}

\usepackage{hyperref}
\hypersetup{
    colorlinks=true,
    linkcolor=blue,
    filecolor=magenta,      
    urlcolor=cyan,
    citecolor=blue,
    pdftitle={Overleaf Example},
    pdfpagemode=FullScreen,
    }

\makeatletter
\renewcommand*\aa@pageof{, page \thepage{} of \pageref*{LastPage}}
\makeatother

\begin{document}

    \title{Probing chromospheric fine structures with a H$\alpha$ proxy using MURaM-ChE}

    \author{Sanghita Chandra\inst{1},          
          Robert Cameron\inst{1},
          Damien Przybylski\inst{1}, 
          Sami K. Solanki\inst{1},
          Patrick Ondratschek,\inst{1} \\
          \and 
          Sanja Danilovic\inst{2}
          }

   \institute{Max Planck Institute for Solar System Research,
              Justus von Liebig Weg, 37077 G\"ottingen, Germany\\
              \email{chandra@mps.mpg.de}
        \and
            Institute for Solar Physics, Dept. of Astronomy, Stockholm University, Albanova University Center, 10691 Stockholm, Sweden
             }

   \date{Received: 23 May 2025 / Accepted: 06 August 2025}

 
\abstract
{H$\alpha$ observations of the solar chromosphere reveal dynamic small-scale structures known as spicules at the limb and rapid blueshifted and redshifted excursions (RBEs and RREs) on-disc. 
}
{We want to understand what drives these dynamic features, their magnetohydrodynamic (MHD) properties, and their role in energy and heat transport to the upper solar atmosphere. To do this, we aim to develop a proxy for synthetic H$\alpha$ observations in radiative-MHD simulations to help identify these features.}
{We used the chromospheric extension to the MURaM code (MURaM-ChE) to simulate an enhanced network region. We developed a proxy for H$\alpha$ based on a photon escape probability. This is a Doppler-shifted proxy that we used to identify fine structures in the line wings. We studied on-disc features in 3D, obtaining their 3D structure from the absorption coefficient.}
{We validate the H$\alpha$ proxy by comparing it against features detected in the wings of H$\alpha$ synthesised using MULTI3D. We detect numerous small-scale structures rooted at the network patches, similar to observations in H$\alpha$. The dynamics of an example feature (RBE) at a Doppler shift of 37 km/s show that flux emergence and consequent reconnection drive the formation of this feature. Pressure gradient forces build up to drive a flow along the field line carrying the feature, making it a jet. There is strong viscous and resistive heating on the first appearance of the feature associated with the flux emergence. At the same time and location, a heating front appears and propagates along the field lines at speeds comparable to the  Alfv\'en velocity. The feature shows an oscillatory behaviour as it evolves. }
{We show that a synthetic observable based on an escape probability is able to reliably identify features observed with the H$\alpha$ spectral line. We demonstrate its applicability by studying the formation, dynamics and properties of an RBE.}


    \titlerunning{Chromospheric H$\alpha$ structures}
    \authorrunning{Chandra et al.}

   \maketitle
%

\section{Introduction}


   \begin{figure*}
   \centering
   \includegraphics{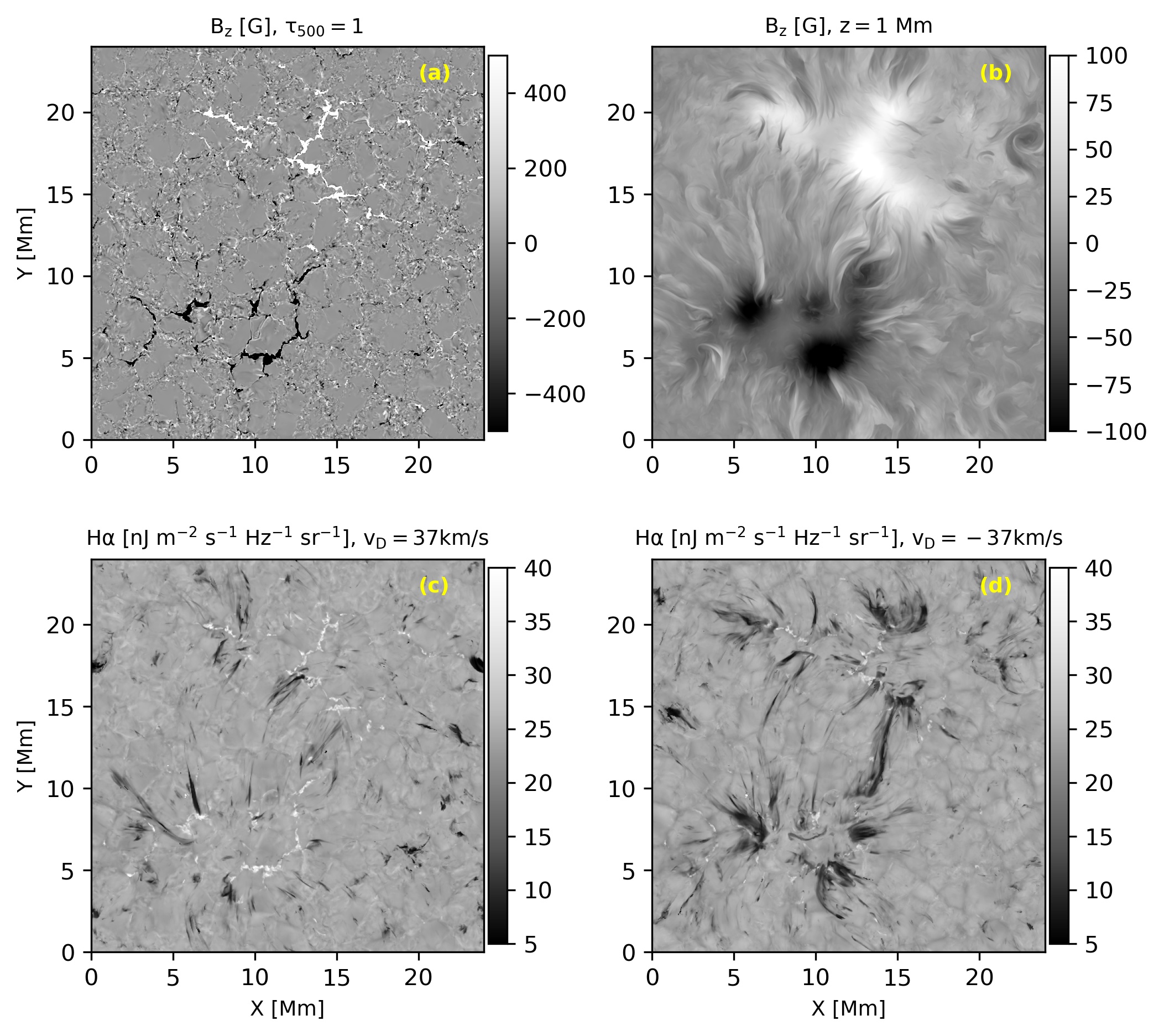}
   \caption{Overview of the magnetic field and synthetic H$\alpha$ intensity in a snapshot of the enhanced network simulation. In panels (a) and (b) we show the vertical component of the magnetic field ($\mathrm{B_z}$) at the surface ($\tau_{500}$ = 1) and in the chromosphere (z = 1 Mm), respectively. Panels (c) and (d) are synthetic H$\alpha$ intensities in the blue (37 km/s) wing and red (-37 km/s) wing, respectively. $\mathsf{v}_\mathrm{D}$ in the title of panels (c) and (d) denotes the Doppler shift.}
              \label{overview}%
    \end{figure*}

   \begin{figure}
    \includegraphics[width=0.5\textwidth]{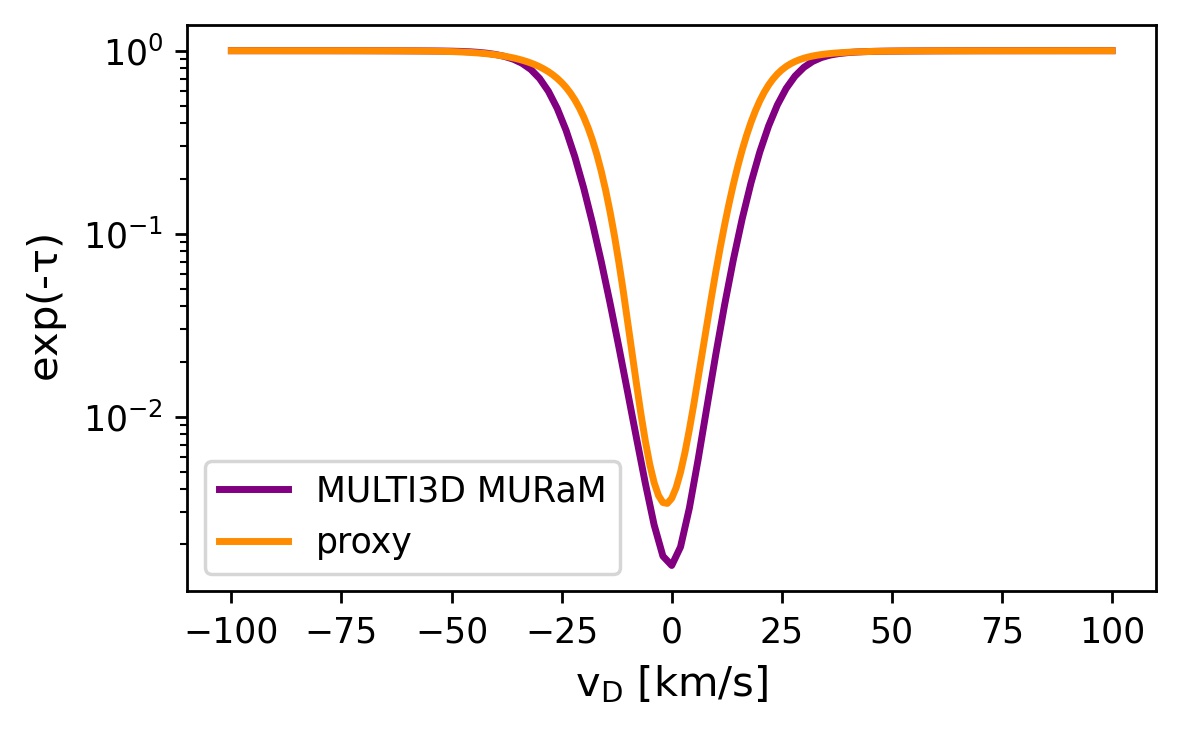}
   \caption{Spatially averaged profile of the H$\alpha$ proxy, modelled as an escape probability (orange). The same quantity from the MULTI3D MURaM (-ChE) synthesis for comparison (purple). }
              \label{Compare_synthesis_proxy}%
    \end{figure}
    
   \begin{figure}
        \includegraphics[width=0.5\textwidth]{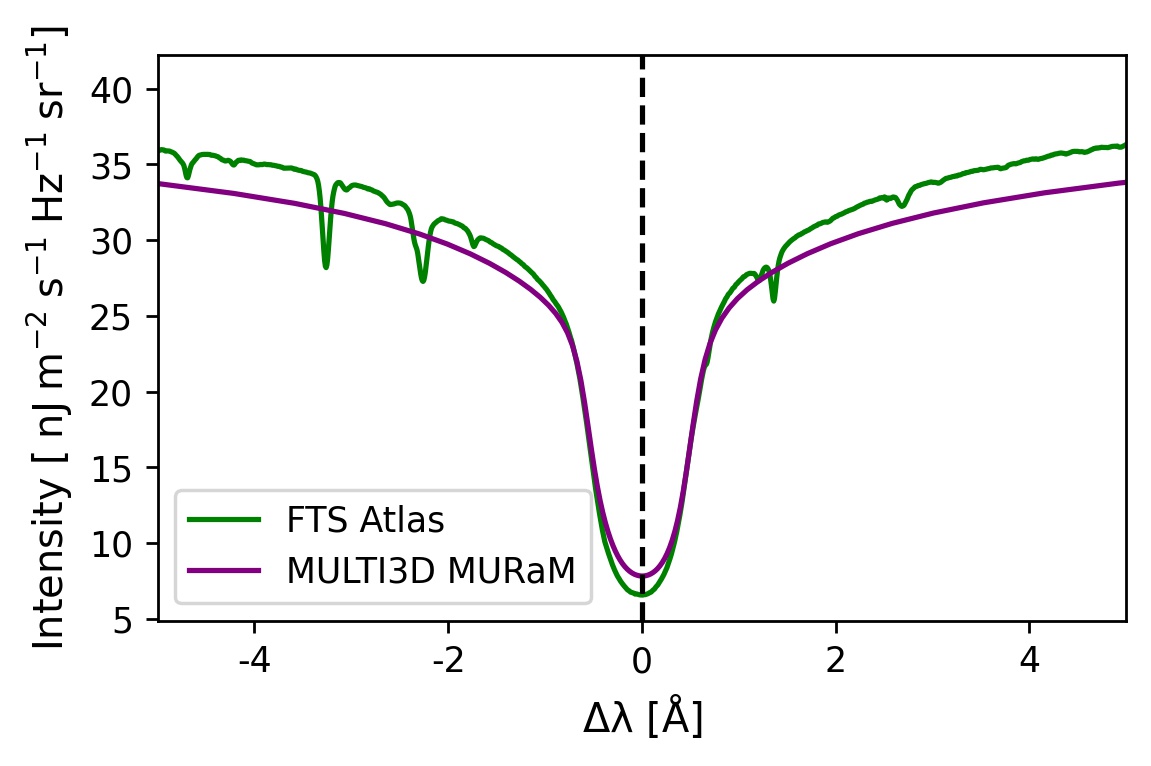}
        \caption{Spatially averaged H$\alpha$ line profile from the FTS Atlas of \cite{Neckel_1984} (green) compared with the H$\alpha$ profile synthesised by applying MULTI3D to the simulated MURaM-ChE atmosphere (purple). The dashed black line marks the centre of the line core. We have shifted the synthetic profile (MULTI3D MURaM) by 0.035 $\AA$ ($\approx$ 1.59 $\mathrm{km/s}$)} towards the blue wind to align the line cores for comparison.
        \label{Benchmark}
    \end{figure}

    \begin{figure}
    \includegraphics[width=0.5\textwidth]{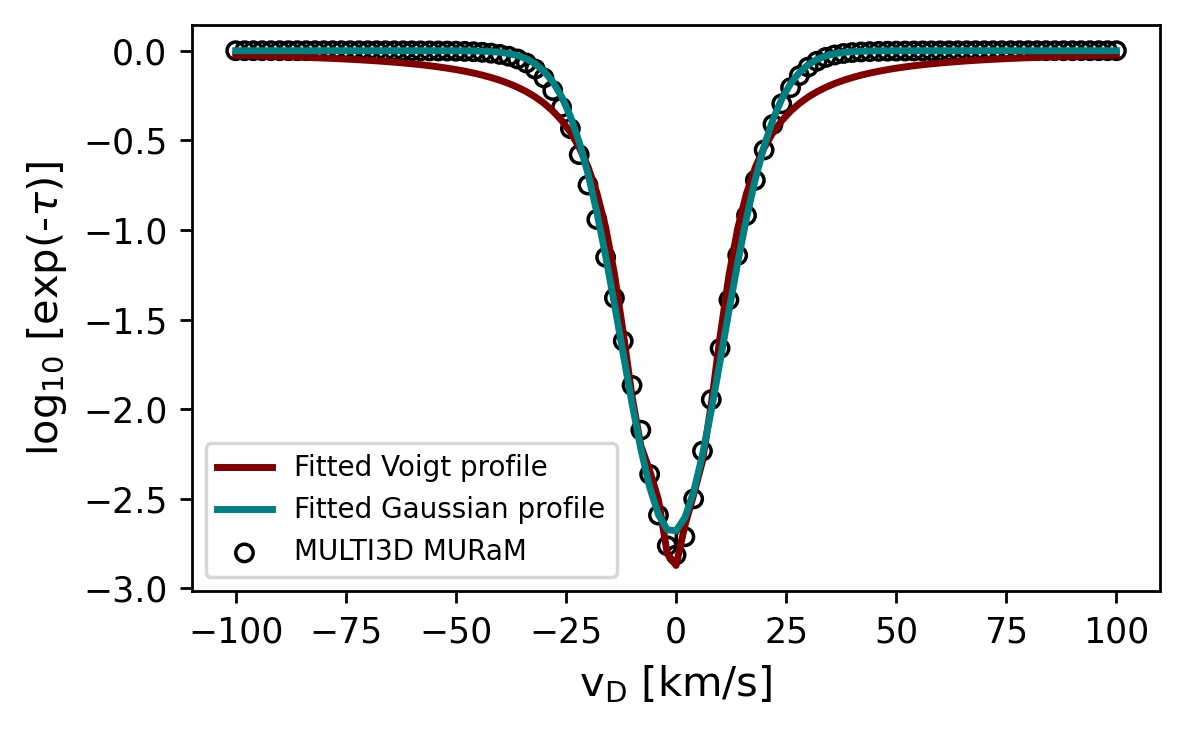}
   \caption{Average escape probability profile from the MULTI3D synthesis on MURaM-ChE (black circles) fitted with a Voigt profile (maroon) and a Gaussian profile (teal).}
              \label{Fitting}%
    \end{figure}

   \begin{figure*}
   \centering
   \includegraphics[width=1\textwidth]{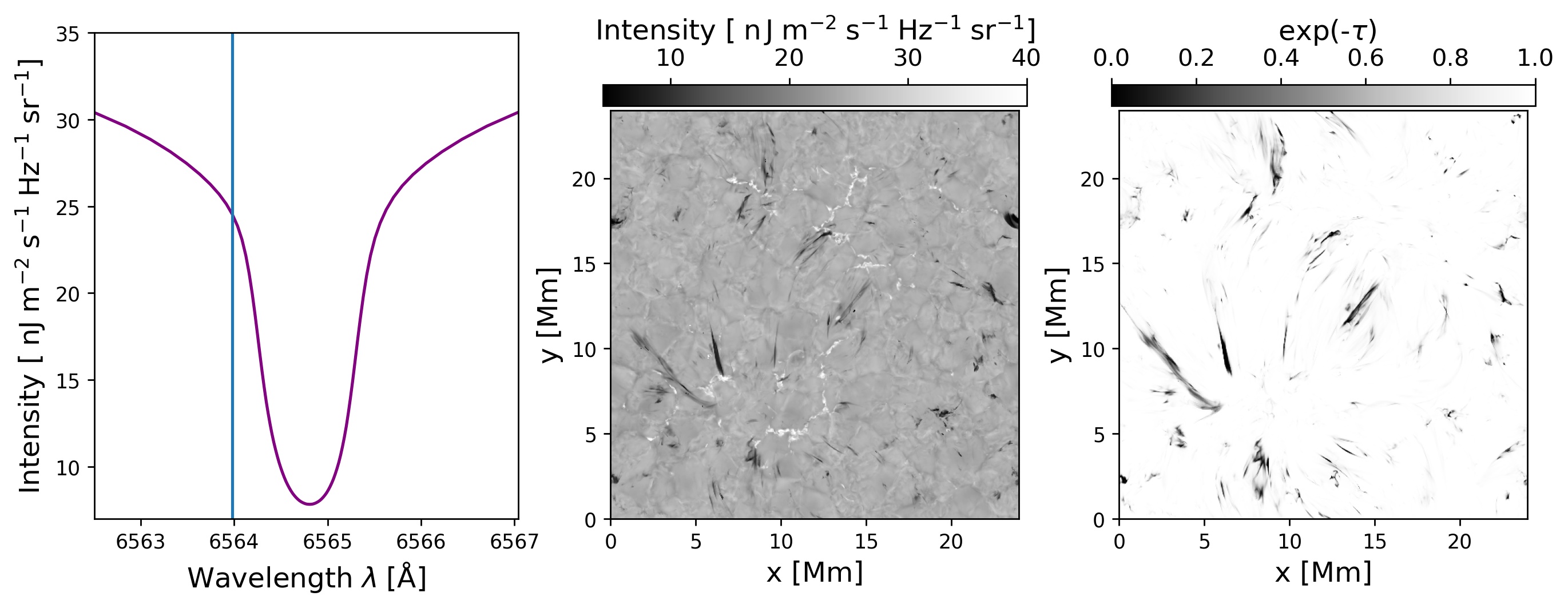}
   \caption{Features in the blue wing at a Doppler velocity of 37 km/s. The leftmost panel shows the average line profile for the synthesised H$\alpha$ intensity (using MULTI3D) with wavelength. The vertical blue line marks the wavelength at which the Doppler velocity is 37 km/s with respect to the line centre ($\mathrm{\Delta \lambda} = 0$). The middle panel shows the synthetic H$\alpha$ intensity at this wavelength. The third panel shows the proxy at the same wavelength. A scan through the line is available \href{https://owncloud.gwdg.de/index.php/s/3QUzqzqGjgBmewg}{online}.}
              \label{Bluewing}%
    \end{figure*}

The solar chromosphere is a highly dynamic and structured layer of the Sun’s atmosphere, bridging the gas pressure-dominated photosphere and the magnetically dominated corona. Its intricate interplay of radiation, magnetic fields, and plasma makes it a challenging region to study \citep{Carlsson_2019}. Among its most striking features are spicules, rapid blueshifted excursions (RBEs), and rapid redshifted excursions (RREs), which potentially play a crucial role in transporting mass and energy to the upper solar atmosphere.

A key tool for studying chromospheric dynamics is the H$\alpha$ line, one of the most widely used spectral lines for solar observations \citep{Rutten_2007, Rutten_2008}. Its strong absorption features on the solar disc and emission features near the limb make it well suited for imaging and spectroscopic techniques. The morphology and evolution of several large features such as filaments and prominences are  best observed in H$\alpha$. In its wings, H$\alpha$ works well to detect fine structures such as spicules on much smaller length scales --- down to several hundred kilometres  \citep{secchi_1871,Beckers_1968, Beckers_1972, Pikelner_1971}.

Spicules appear as elongated, fine multi-threaded strands that abound in the solar chromosphere at the limb. They are easily observed in the wings of the H$\alpha$ and Ca II lines \citep{de_Pontieu_2007b, Sykora_2009}. These dynamic features invite interest because their ubiquity and dynamic motion make them good candidates for mass and heat transport to the upper solar atmosphere. 
Spicules are categorised into types I and II. Type I spicules are observed to move up and fall back down with apparent velocities of 15-40 km/s in height-time plots, tracing a parabolic trajectory \citep{Rouppe_van_der_Voort_2009, Pereira_2012}. They exhibit lifetimes of 3-10 minutes and are driven by magneto-acoustic shocks \citep{de_Pontieu_2004}. The on-disc counterparts of type I spicules are the dynamic fibrils. Type II spicules are more dynamic, showing rapid evolution with lifetimes of a few seconds up to 4 minutes. Some of the type II spicules are seen to disappear with high velocities (80-300 km/s), with nothing falling back to the surface after the rising phase. The on-disc counterparts of type II spicules are identified as RBEs (\citealp[and others]{Langangen_2008,Rouppe_van_der_Voort_2009,Sekse_2012, Shetye_2016}). The RREs share similar morphologies and statistical properties to the RBEs, appearing in the red wing of H$\alpha$ and Ca II lines. The RREs could be return flows of type II spicules from the transition region \citep{Chaurasiya_2024}.

The rapid excursions (RBEs and RREs) are concentrated around the magnetic network and plage, with their footpoints rooted near strong magnetic field concentrations \citep{Bose_2021, Danilovic_2023}. First observed in the Ca II IR line \citep{Langangen_2008}, they are characterised by lifetimes of up to a few minutes, linear trajectories, upflow speeds for RBEs (20–50 km/s), and distinct lengths (up to $\sim$ 6 Mm) and widths (a few hundred kilometres) in observations. Although we have a grasp on the statistical properties of type II spicules, RBEs and RREs, our understanding of their driving mechanisms and energetics remains incomplete \citep{Pereira_2012, Sekse_2013}.

Spicules are particularly intriguing due to the velocities that govern their motion. Apparent velocities of type II spicules can be exaggerated by line-of-sight density effects, while actual plasma motions may be much slower. The use of Doppler velocities further complicates the interpretation, as rapid flow direction changes can cause features to disappear from one spectral passband and reappear in another \citep{Pereira_2016}. The complex transverse and torsional motions observed in rapid excursions \citep{Sekse_2013} also mean that RBEs can switch to RREs and vice versa, depending on the line of sight and the direction of plasma flows.

Simulations are an important tool to help us interpret the observations. Synthetic H$\alpha$ images can be generated from the simulations to look for similar events to those which are observed. 
In simulations we have the  full 3D structure which is important for understanding the dynamics.
One approach to synthesising H$\alpha$, given a background atmosphere, is to assume that the hydrogen population levels are in statistical equilibrium with the non-local thermodynamic equilibrium (NLTE) radiation field \citep{Leenaarts_2012}. This implicitly assumes that the NLTE radiation field and populations reach equilibrium on timescales shorter than those of the dynamics.
The resulting problem then involves solving simultaneously for the radiation field and populations: this is the approach adopted by codes such as RH and MULTI3D \citep{Uitenbroek_2001, Leenaarts_and_Carlsson}.

An alternative approach is to use the populations calculated by MURaM-ChE using the rate equations \citep{Przybylski_2022}. The rate equations implemented in MURaM-ChE include the dynamical effects but include only a very approximate treatment of the NLTE radiation field \citep{Sollum_1999}. These MURaM-ChE populations can be used in the H$\alpha$ synthesis in several ways with different levels of sophistication. One approach investigated by \cite{Krikova_2023} is to synthesise H$\varepsilon$ using a modified version of the RH code that includes the non-equilibrium hydrogen ionisation from a Bifrost simulation. In this paper we introduce the simplest way. We show that our approach with MURaM-ChE and MULTI3D synthesis (which assumes instantaneous equilibrium between the populations and the NLTE radiation field) leads to the identification of the same features, which is sufficient for our study.

Numerical modelling of solar spicules dates back to the late 1960s (see the review by \citealp[]{Sterling_2000}). Multiple models describing the driving mechanism of spicules have been proposed such as Alfv\'en waves, pressure pulses in the high chromosphere, shocks, etc. More recent simulations performed by \cite{Sykora_2017} outline the role of ambipolar diffusion in generating a solar spicule via the build-up of magnetic tension. There have been conjectures regarding propagating heating events (PHEs) associated with the formation of type-II spicules, RBEs and RREs \citep{Rutten_2017}. Using H$\alpha$ observations from the Goode Solar telescope at the Big Bear Solar Observatory \cite{Samanta_2019} show evidence of magnetic reconnection driving spicules and subsequent coronal heating. Using radiative-magnetohydrodynamic (radiative-MHD) simulations \cite{Druett_2022} demonstrate how the destabilisation of low-lying twisted magnetic field lines and Lorentz forces contribute to the formation, mass-loading, and drainage of dynamic fibrils. \cite{Danilovic_2023} compute the H$\alpha$ line in 3D radiative-MHD simulations to reproduce RBEs and RREs and assert that Alfv\'enic waves are the primary driver. Studies also suggest that the apparent motions of these jet-like features correspond to propagating heating fronts rather than actual mass motions \citep{de_Pontieu_2007b}. In a study by \cite{Bose_2021}, the heating in downflowing RREs is ascribed to ambipolar diffusion \citep{Sykora_2017}.

Synthetic observables are crucial for interpreting spicules, RBEs, and RREs in simulations. There have only been three attempts aimed at modelling RBEs and RREs: \cite{Kuridze_2015}, \cite{Srivastava_2017}, and \cite{Danilovic_2023}. These studies also show that it is not straightforward to model these on-disc fine structures such that they reproduce properties (e.g. lengths, lifetimes, and motion) as seen in observations. For our purpose, we develop a synthetic observable based on an escape probability \citep{Hummer_1982} for the H$\alpha$ spectral line.

In this work we introduce a Doppler-shifted H$\alpha$ proxy (Sect.\,\ref{Proxy}), which was computed using the  populations of the different levels of hydrogen calculated in a 3D MURaM-ChE simulation. 
These populations were calculated using the rate equations and include information from the history of the plasma. 
This is a different approach from assuming instantaneous  statistical equilibrium which is usually assumed in radiative transfer models such as RH and MULTI3D. Because we do not solve for the population levels, under the assumption of statistical equilibrium with the radiation field, the proxy synthesis is computationally cheap (a few CPU seconds for each snapshot) and can be done on the fly with MURaM-ChE with almost no additional costs. Despite the differences in the two approaches, we get very similar results in the morphology of on-disc H$\alpha$ wing features for both the proxy and the MULTI3D synthesis for the two snapshots we study. 
The proxy acts as a tool for identifying fine-structures that can then be studied in various ways. Here we concentrate on a proof of concept based on a detailed study of a single feature. A statistical investigation of many such features in a MURaM-ChE simulation will be presented in an upcoming paper.

 In this study we focus on the driving mechanism of an example RBE feature and associated heating. The properties of the feature are evaluated and compared to those of a typical RBE, as seen in observations (see Sect. \ref{feature-of-interest}). Section \ref{driving_mechanism} focuses on the forces and flows driving the feature during its lifetime. The identified feature exhibits jet-like behaviour. The origin of this jet is discussed. We also observe an oscillatory behaviour during its evolution. Finally, we provide further discussions and the conclusions in Sect. \ref{discussion}. 

\section{Model atmosphere and synthesis}\label{model_atmosphere}

    \subsection{Enhanced network model with MURaM-ChE}

    The MURaM radiative-MHD code \citep{Voegler_2005, Rempel_2017} models the near-surface convection, photosphere, chromosphere, and corona in 3D Cartesian geometry. With the chromospheric extension to MURaM --- hereby MURaM-ChE \citep{Przybylski_2022}, the code now includes an equation of state that treats hydrogen level populations in non-equilibrium (NE). This treatment involves solving the NE rate equations for hydrogen and is essential to forward model the formation of H$\alpha$.

    A simulation of an enhanced network region, performed using MURaM-ChE is used for this study. The grid dimensions are $24\,\mathrm{Mm}\:\times\:24\,\mathrm{Mm}\:\times24\,\mathrm{Mm}$ with a resolution of $23.46\,\mathrm{km}\:\times23.46\,\mathrm{km}\:\times20\,\mathrm{km}$ in the x, y and z (vertical) directions, respectively. The simulation spans from approximately $7~\mathrm{Mm}$ below the photosphere (at z $\approx$ 0\,Mm), to $17~\mathrm{Mm}$ above. It has a bipolar magnetic field configuration aimed to model a network field patch on the Sun (the same model is described in greater detail in \citealp{Ondratschek_2024}). We used 101 snapshots spanning 606 s, with a cadence of $\sim$ 6 s, in order to perform the analysis. An overview of the magnetic field at different heights is shown in panels (a) and (b) of Fig \ref{overview}. An enhanced magnetic network with the bipolar configuration is clearly visible at the photosphere, and it extends into the chromosphere. In panels (c) and (d) of Fig. \ref{overview} we show synthetic H$\alpha$ intensity maps generated using MULTI3D (see Sect. \ref{synHalpha}). 

    \subsection{Synthesis with MULTI3D}\label{synHalpha}
     We used MULTI3D \citep{Leenaarts_and_Carlsson} with the modelled atmosphere to forward model the H$\alpha$ spectral line. The MULTI3D code solves the radiation transfer problem in NLTE on a 3D Cartesian grid. We treated the H$\alpha$ line under the approximation of complete-frequency-redistribution (CRD). The viewing angle was set to $\mu$ = 1, to study on-disc features. Every second column was sampled in the two horizontal directions to make the computation faster (see for example \citealp{Danilovic_2023}). The vertical height limit for this computation was set from -1 Mm to 11 Mm. Most of the cool chromospheric material lies below 10 Mm. This determined our upper limit for the synthesis. The computation for H$\alpha$ was done using a five-level plus continuum hydrogen model atom \citep{Leenaarts_2012}. MULTI3D solves the RT equation over multiple frequency points (order of $10^2$ points), with the assumption of statistical equilibrium.

\section{Identifying fine structures in H$\alpha$}\label{Proxy}
We focus on identifying features in the wings of the average H$\alpha$ spectral line within the domain where H$\alpha$ is synthesised. We accounted for the NLTE physics essential for H$\alpha$ formation using MURaM-ChE. We then used MULTI3D to forward model the H$\alpha$ line from the MURaM-ChE atmosphere. This process is computationally expensive ($\sim$ $10^5$ CPU hours per snapshot). To study temporal evolution efficiently, we employed a faster approximation that generates proxy H$\alpha$ maps in seconds using NE hydrogen populations, enabling feature identification in the line wings.

\subsection{Description of the proxy: Two-level hydrogen model atom}\label{proxy_description}

The H$\alpha$ proxy is designed to be modelled based on the probability of photon escape from the photosphere and chromosphere. On the solar disc H$\alpha$ is in absorption (n=2 $\rightarrow$ n=3 transition). MURaM-ChE computes the hydrogen populations for a five-level plus continuum hydrogen model atom. In our fast method we used a two-level approximation (involving energy levels: n=2, n=3) of the hydrogen atom.  We also assumed uniform, time-independent radiation coming from the photosphere. The simulated atmospheric parameters --- temperature and velocity --- were utilised to estimate the line profile function ($\Phi$). Since the H$\alpha$ line is broadened predominantly by the small-scale thermal and turbulent motions (Doppler broadening) on the Sun, $\Phi$ takes the form

   \begin{align} \label{line_profile}
         & & {\Phi(\mathsf{v}_\mathrm{D}) = \frac{c}{\nu_0}\sqrt{\frac{m}{ {2\pi k_\mathrm{B} T}} }  \mathrm{exp}\left(-\frac{m(\mathsf{v}-\mathsf{v}_\mathrm{D})^2}{2k_\mathrm{B}T}\right)},
   \end{align}

 \noindent  where `$\mathsf{v}$' is the line-of-sight velocity ($\mathsf{v_z}$ in our case). $\mathsf{v}_\mathrm{D}$ is the Doppler velocity. \textit{T} is the temperature, \textit{m} is the atomic mass of hydrogen, and $k_\mathrm{B}$ is the Boltzmann constant.
\newline For H$\alpha$ we took the wavelength ($\lambda_0$) of 656.28 nm at the line core (in air), which corresponds to a frequency, $\nu_0$. Following \cite{Rybicki_1979} we computed a measure for the opacity in the H$\alpha$ line from the absorption coefficient ($\mathrm{\kappa}$), defined as 

   \begin{align}\label{abs_coeff}
      & & \kappa = \frac{h \nu_0}{4 \pi} (n_2 B_{23} - n_3 B_{32}) \  \Phi(\mathsf{v}_\mathrm{D}) ,
   \end{align}

\noindent where $n_2$, $n_3$ correspond to hydrogen populations in the (n=2) and (n=3) energy levels, respectively. $B_{23}$, $B_{32}$ are the Einstein coefficients. The first component ($n_2 B_{23}$) corresponds to absorption, while the second term (${n_3 B_{32}}$) corresponds to stimulated emission. We have the following Einstein's relations:

   \begin{align}
     & &  \frac{A_{32}}{B_{32}} = \frac{2 h \nu_0^3}{c^2}, &   & \frac{B_{23}}{B_{32}} = \frac{g_3}{g_2}.
   \end{align}\label{einstein_relation}


    \begin{figure*}
   \centering
   \includegraphics[width=0.95\textwidth]{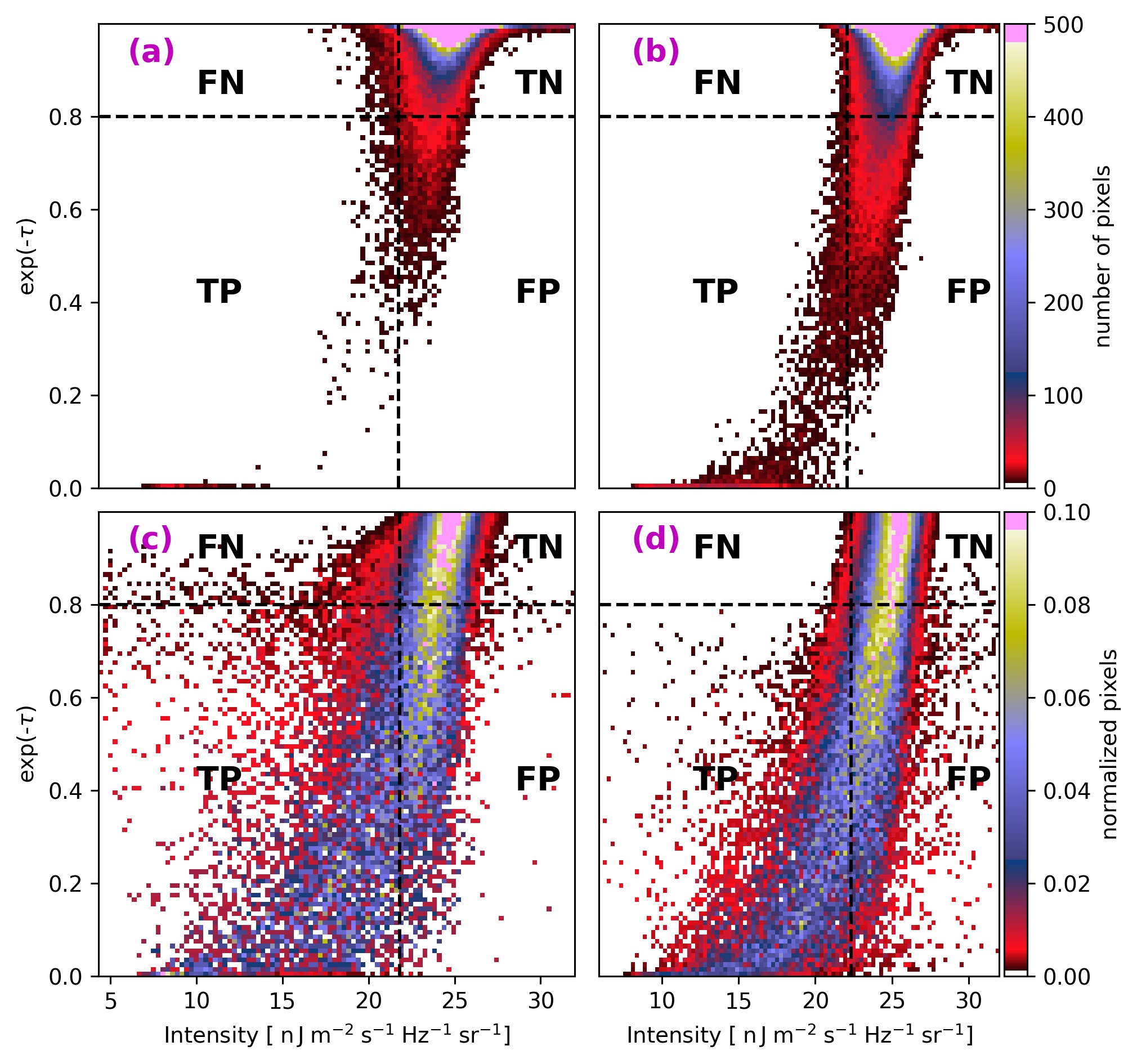}
   \caption{2D histogram plots showing the classification of pixels. Panels (a), (b): 2D histograms of the proxy vs. the synthetic H$\alpha$ intensity at Doppler velocity = 37 km/s (a) and -37 km/s (b). Panels (c), (d): The row normalised distribution of the same pixels corresponding to panels (a) and (b), respectively. We mark the dynamic threshold for the synthetic intensity with vertical dashed black lines. The static threshold for the proxy for feature identification is marked by the horizontal dashed black lines. The four pixel categories: true positive (TP), true negative (TN), false positive (FP), and false negative (FN) are also indicated on each of the panels.}
              \label{2D_intensity_P3}%
    \end{figure*}

\begin{figure*}[h]
   \centering
   \includegraphics[width=0.95\textwidth]{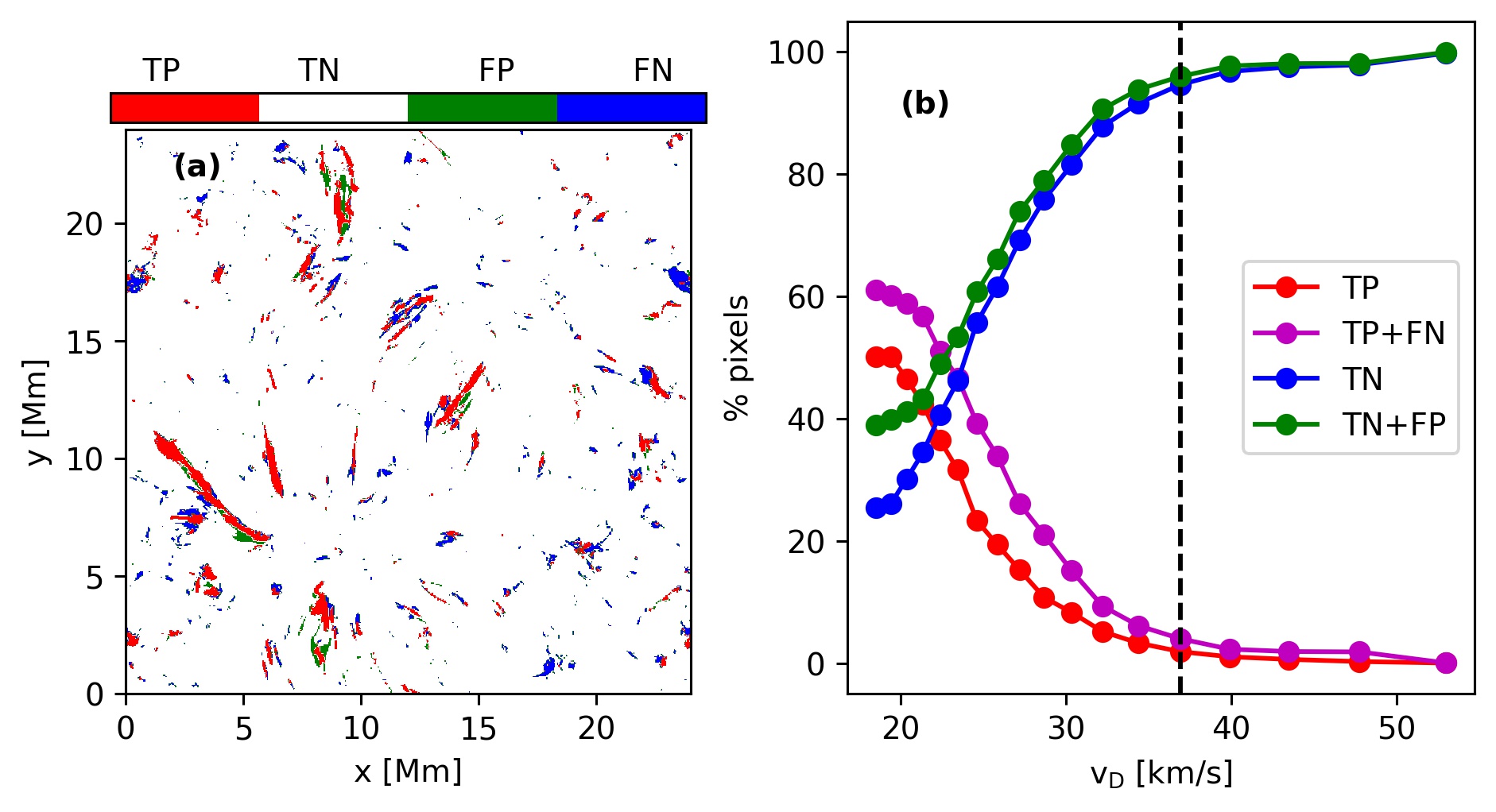}
   \caption{Goodness of the proxy at Doppler velocity = 37 km/s. Panel (a) gives a spatial distribution of (in)correctly identified pixels. The pixels are colour-coded to show true positives (TPs:  red), true negatives (TNs: white), false positives (FPs: green) and false negatives (FNs: blue). Panel (b) represents the percentage of (in)correctly identified pixels as a function of the Doppler velocity ($\mathsf{v}_\mathrm{D}$) in the blue wing of the line. The dashed black line marks the position in the line wing ($\mathsf{v}_\mathrm{D}$ = 37 km/s) at which panel (a) is obtained. A scan through the blue wing of the line exhibiting the goodness of the proxy is available \href{https://owncloud.gwdg.de/index.php/s/asyj1VRKNizSPwJ}{online}. 
   }  
              \label{goodness_499}%
\end{figure*}


For H$\alpha$, the Einstein's coefficient ${A_{32}}$ = $4.41 \times 10^7$ $\mathrm{s^{-1}}$, ${g_3} = 18$ and ${g_2} = 8$ (see NIST\footnote{\url{https://physics.nist.gov/PhysRefData/ASD/lines_form.html}}). Plugging the computed Einstein's coefficients in Eq. (\ref{abs_coeff}), we obtained the absorption coefficient. We then calculated the optical depth for H$\alpha$. Under the assumption that there is no scattering, the absorption coefficient equals the extinction coefficient. Note that while this assumption holds for the line wings, the line core is dominated by scattering. The optical depth for a distance, \textit{z},  is given by

   \begin{align}\label{tau}
      & & \tau(z) = \int_{z = z_1}^{z =z_2} \kappa dz ,
   \end{align}

\noindent where \textit{dz} is the length element in the line of sight. \textit{$z_1, z_2$} denote the vertical extent (top to bottom) over which the integration is performed. The expression \textit{$\kappa$dz} gives the probability that a photon will be absorbed when travelling through distance \textit{dz}. Let \textit{P(z)} be the probability of a photon traversing a distance \textit{z} without getting absorbed. Then it can be shown that

\begin{align}
    & & P(z) = \mathrm{exp}(-\tau).
\end{align}

This treats the absorption in H$\alpha$ as a photon escape probability. For our simulation box, \textit{dz} = 20 km. We chose \textit{${z_1}$} = 6\,Mm and \textit{${z_2}$} = 0.7\,Mm in Eq. (\ref{tau}) to cover heights where most of the fine-structures observed in H$\alpha$ are formed \citep{White_1966}. We used 6 Mm as an upper limit because integrating the atmosphere to higher heights does not affect our results. The average height of the temperature minimum lies close to 0.7 Mm. The average Doppler profile for this proxy is shown in Fig. \ref{Compare_synthesis_proxy}. A shift of the proxy from $\Delta \mathsf{v} = 0$ is visible in this figure. This shift varies in time with an approximate 3-minute period (based on a 10-minute time sequence).

In Fig. \ref{Compare_synthesis_proxy} we also show a comparison using the computed opacities ($\tau$) from MULTI3D for the synthetic H$\alpha$ spectral line to construct the photon escape probability, exp(-$\tau$). The differences between the two escape probability profiles stem from several factors. MURaM-ChE accounts for the non-equilibrium hydrogen populations but assumes a very simple treatment of the radiation field \citep{Sollum_1999}. MULTI3D assumes statistical equilibrium for computing the hydrogen populations but has a more detailed treatment of the RT problem.   

The proxy works as a physically motivated tool. It helps isolate H$\alpha$ absorption features in the simulation domain based on the hydrogen populations at the n=2 level (for absorption), temperature, and the line-of-sight velocities.

\subsection{The synthetic H$\alpha$ intensity maps}

   To test the performance of the proxy against 3D NLTE synthesis, we used MULTI3D to generate synthetic H$\alpha$ intensity maps over a range of wavelengths --- 6543.5$\AA$ to 6591.3$\AA$. Two snapshots were synthesised with the computation described in Sect. \ref{synHalpha}; they are 12 seconds apart. Throughout Sect. \ref{Proxy} we show the analysis using only one snapshot. The results remain unchanged for the other snapshot.
      
    We find numerous small-scale structures in the generated synthetic H$\alpha$ maps (see panels c and d in Fig. \ref{overview}) that appear to be rooted close to the network patches. The SST (Swedish Solar telescope) dataset observed on 25 May 2017 at 09:16:58 UTC shows such an enhanced network region (see for example Fig. 1 in \citealp{Bose_2021}) with numerous RBEs and RREs. Our synthesised H$\alpha$ maps appear to be similar to their observational counterparts.  

   We further compare the observed average H$\alpha$ profile from the FTS Atlas of \citet{Neckel_1984} to the synthetic intensity average profile from our MURaM-ChE simulation (see Fig. \ref{Benchmark}). This helps to benchmark the synthetic H$\alpha$ maps created using MULTI3D-MURaM against observations. The FTS Atlas observations correspond to quiet Sun regions spatially (and temporally) averaged over the disc centre whereas the simulation is that of an enhanced network. Nevertheless, the comparison shows a good match in the line profile, especially in the inner wings. The shift (= 0.035 $\AA$ or 1.59 km/s) that we introduced to align the line core arises from two major sources. We calculated the gravitational redshift to be 0.63 km/s. Secondly, the average velocity at the formation height is a downflow of 0.88 km/s. These factors introduce a slight shift of the line towards the red.

   In the far wings of H$\alpha$, collisional or pressure broadening becomes significant. Consequently, a Voigt profile is generally more suitable for modelling the line than a simple Doppler profile. We fitted the average H$\alpha$ line profile from the MULTI3D synthesis with the best-fit Voigt profile and Gaussian profile (see Fig. \ref{Fitting}). The Gaussian fit aligns more closely with the MULTI3D average profile, while the Voigt profile appears broader in the far wings, as expected. This supports the approximation of using the Doppler profile to generate our proxy. The maximum discrepancy between the Voigt and Gaussian fits is on the order of $10^{-1}$.

\subsection{Results: Goodness of the proxy}

The H$\alpha$ proxy effectively identifies the location and morphology of chromospheric absorption features at $\mu$ = 1 in the line wings (Fig. \ref{Bluewing}) based on the MULTI3D synthesis. However, it becomes less reliable closer to the line core, where numerous feature pixels saturate the maps and the structuring of individual features is no longer visible (see movie associated with Fig. \ref{Bluewing}). This arises because of the simplifying assumptions that went into the construction of the proxy (see section \ref{proxy_description}). Despite the simplifications, our assumptions are sufficient for the scope of this study to capture absorption features in the H$\alpha$ wings.

To quantify the goodness of our proxy, we performed a pixel-by-pixel comparison. This was done by generating 2D intensity distributions for the synthetic MULTI3D image and the proxy (see Fig. \ref{2D_intensity_P3}). We find a positive correlation in the pixel distribution shown in Fig. \ref{2D_intensity_P3}. This means lower synthetic intensities (with MULTI3D) correspond to lower photon escape probabilities in the proxy and vice versa. In the wings there are fewer dark absorption features and these occupy a smaller area compared to the background intensity coming from the photosphere. Therefore, the major concentration of the pixels (bright pink patch on the top right of the 2D intensity distributions in panels a and b in Fig. \ref{2D_intensity_P3}) indicate the background or photospheric contribution in the wings. We chose thresholds in the 2D intensity maps for the classification of pixels. The threshold for identifying a feature pixel in the synthetic intensity was determined by the background contribution, such that the pixels corresponding to the background (bright pink patch), were identified as negative pixels. This threshold was determined automatically based on the 2D intensity maps and moves with different positions in the line wings. On the contrary our proxy, being an escape probability, is limited to values from 0 to 1. So, we fix a threshold at 0.8. Any pixel having a value below the thresholds belongs to an H$\alpha$ feature. Based on this we obtain four regions in the 2D intensity maps and classify the pixels as true positives (TPs), true negatives (TNs), false positives (FPs), and false negatives (FNs). The classification of pixels into the four categories is further shown in Table \ref{table:1} and in Fig. \ref{2D_intensity_P3}.

\begin{table}
\caption{Classification of pixels}\label{table:1}
\centering
\begin{tabular}{ |c|c|c| } 
 \hline
 Feature & Synthetic Image & not in Synthetic Image \\
 \hline
 not in Proxy & FN & TN \\ 
 \hline
 Proxy & TP & FP \\ 
 \hline
\end{tabular}
\vspace{0.2cm}
\tablefoot{Positives or (TP+FN) indicate pixels that are a feature, whether identified by the proxy or not. Negatives or (TN+FP) are pixels that are not a feature, whether identified as such by the proxy or not.}
\end{table}

After categorising the pixels for a given map, we quantified the goodness of our proxy. In Fig. \ref{goodness_499} we show the quantification of the goodness at a particular Doppler velocity of $\mathsf{v}_\mathrm{D}$ = 37\,km/s. As we move from the wings towards the line core, the absorption features cover larger areas and the background becomes less prominent. So we have a higher percentage of TPs and fewer TNs closer to the line core and vice versa. From the scan in the blue wing (see movie associated with Fig. \ref{goodness_499}), it is clear that the proxy reasonably identifies blueshifted H$\alpha$ on-disc features. We obtain consistent results in the red wing.

We applied this method to both synthesised snapshots with the calculated synthetic H$\alpha$ intensities (with MULTI3D) and obtain consistent results. Our proxy correctly identifies the H$\alpha$ wing features observed in the synthetic images. It is also computationally efficient as it takes a few CPU seconds to obtain the proxy for each snapshot. Thus, we have a quick and robust proxy which can be used to study features that appear to be RBEs and RREs in H$\alpha$. With this proxy we can locate and identify synthetic observables directly from the MURaM-ChE simulations.

Over and beyond using the line-of-sight integrated proxy maps, we can also apply the proxy to vertical cuts. This is then the normalised absorption coefficient ($\mathrm{\kappa}$). We find that this works well to identify at which heights the feature forms in the atmosphere. In a way this serves as a proxy to the formation height of the identified features. This enables a 3D perception of the feature. In the following section, we show the analysis of a particular blueshifted feature using the devised method.

\begin{figure*}
   \centering
   \includegraphics[width=1\textwidth]{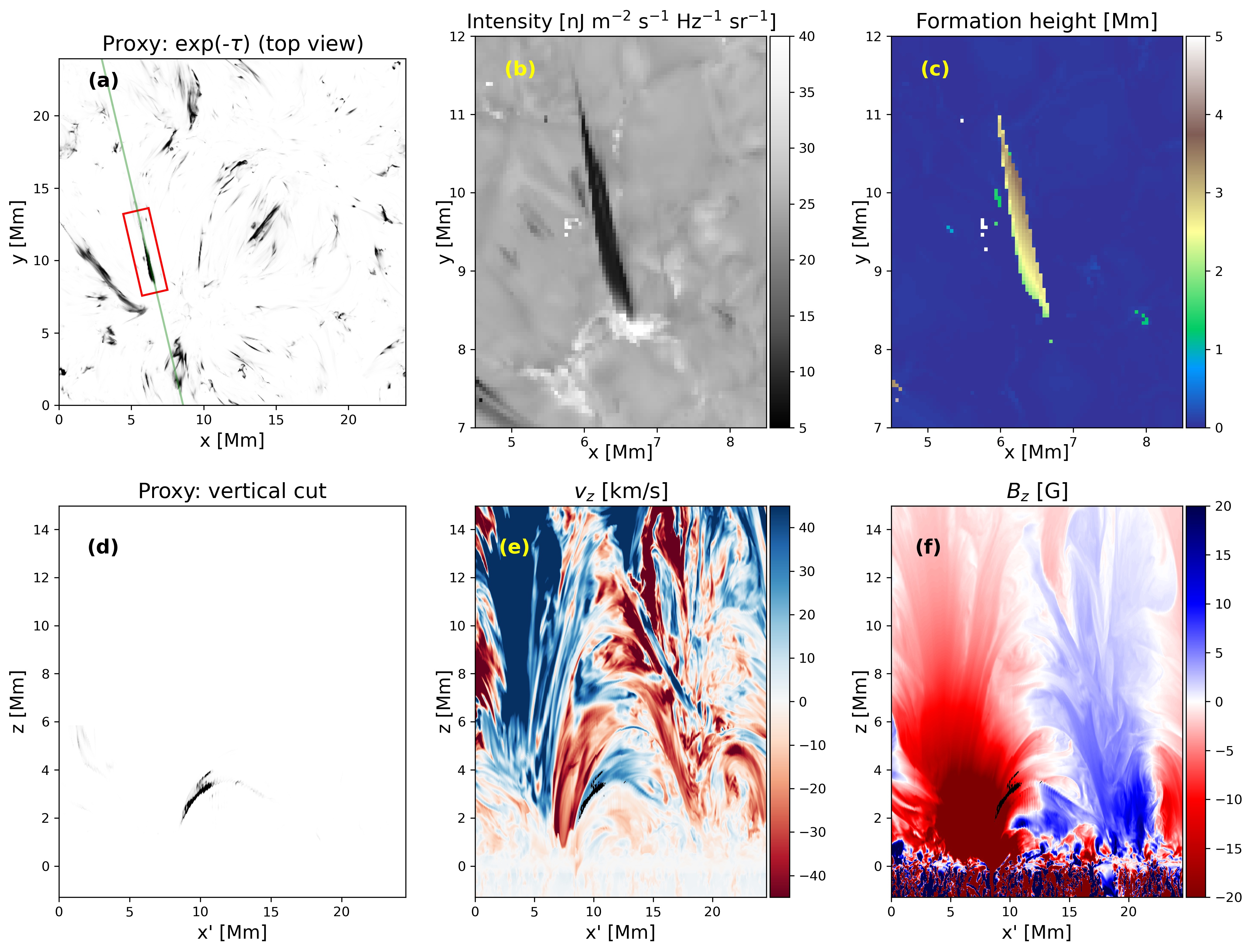}
   \caption{Overview of one blueshifted feature ($\mathsf{v}_\mathrm{D}$ = 37 km/s) detected by the proxy. Panel (a) shows the proxy map with the feature of interest highlighted by the red box. The green line passing through the feature marks the base for the vertical cut through the feature (hereby x' axis). The 0 of x' corresponds to (x,y) = (8.6,0) Mm in the top view (in panel a). Panels (b) and (c) are a zoomed-in view of the feature showing the synthetic intensity in H$\alpha$ and formation height, respectively (obtained with the MULTI3D synthesis). In panel (d) we take a vertical cut through the feature in the proxy (normalised $\kappa$). Panels (e) and (f) show the same vertical cut for the line-of-sight (i.e. vertical) component of the velocity and magnetic field, respectively, with the feature over-plotted in black. A time-lapse of the proxy map in the top view and vertical cuts of various MHD quantities is available \href{https://owncloud.gwdg.de/index.php/s/z9WoEfpvkahEusL}{online}.}
              \label{37F1}%
\end{figure*}

\begin{figure*}
   \centering
   \includegraphics[width=1\textwidth]{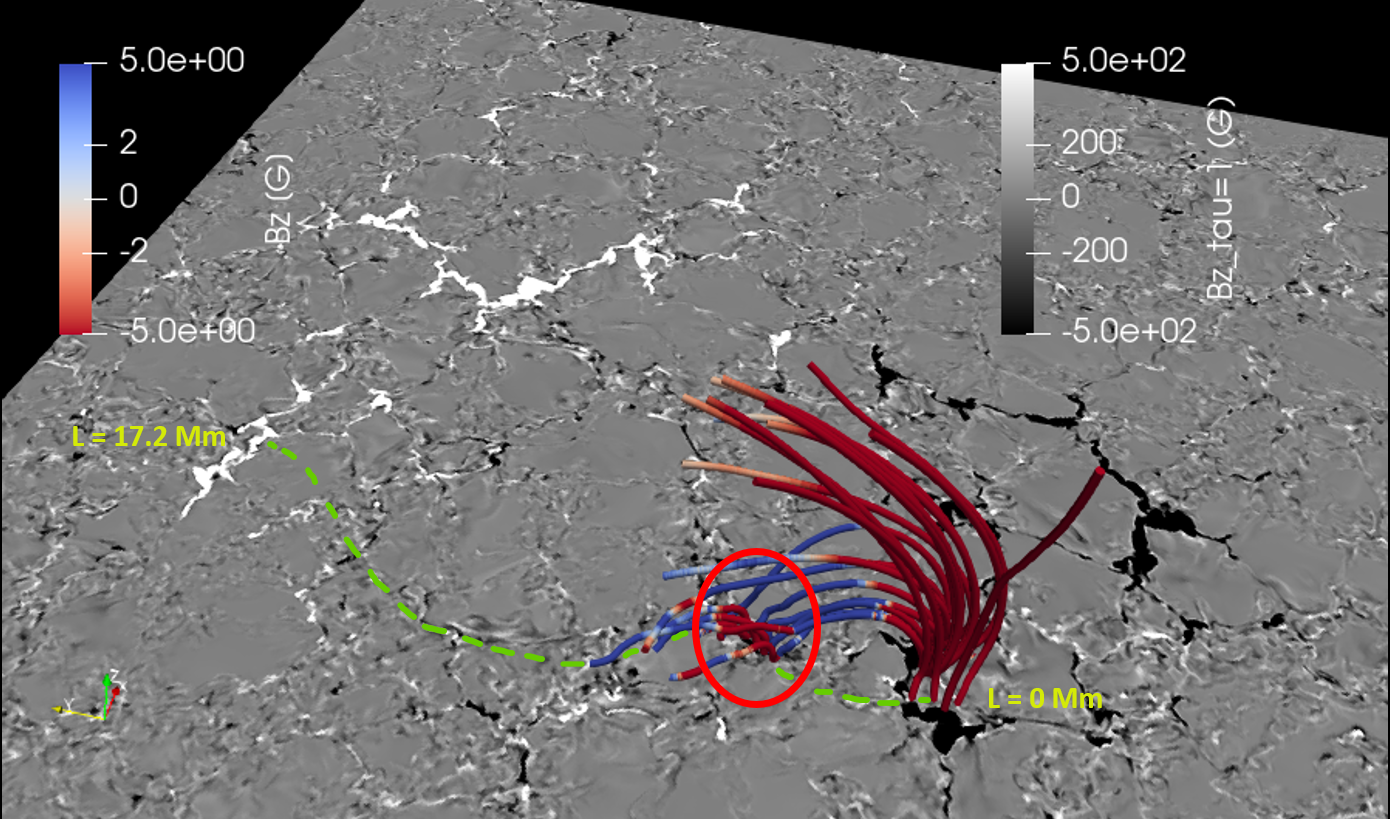}
   \caption{3D rendering of magnetic field lines rooted at the footpoint of the feature at t = 0 s, the first appearance of the feature. The line-of-sight component of the magnetic field is shown as a surface map at the $\tau_{500}$ = 1 surface. The 3D field lines are coloured by the vertical ($\mathrm{B_z}$) component. The site of magnetic reconnection lies across the main footpoint and is marked by the red circle. The dashed green line shows the xy projection of the 3D field line carrying the feature in this snapshot. The magnetic field lines are clipped for better visualisation. A time-series showing the rearrangement of magnetic field lines at the time of feature formation is available \href{https://owncloud.gwdg.de/index.php/s/uJgMYGhgLdtGapY}{online}.}
              \label{bz-topo-before}%
\end{figure*}

\begin{figure}
   \includegraphics[width=0.5\textwidth]{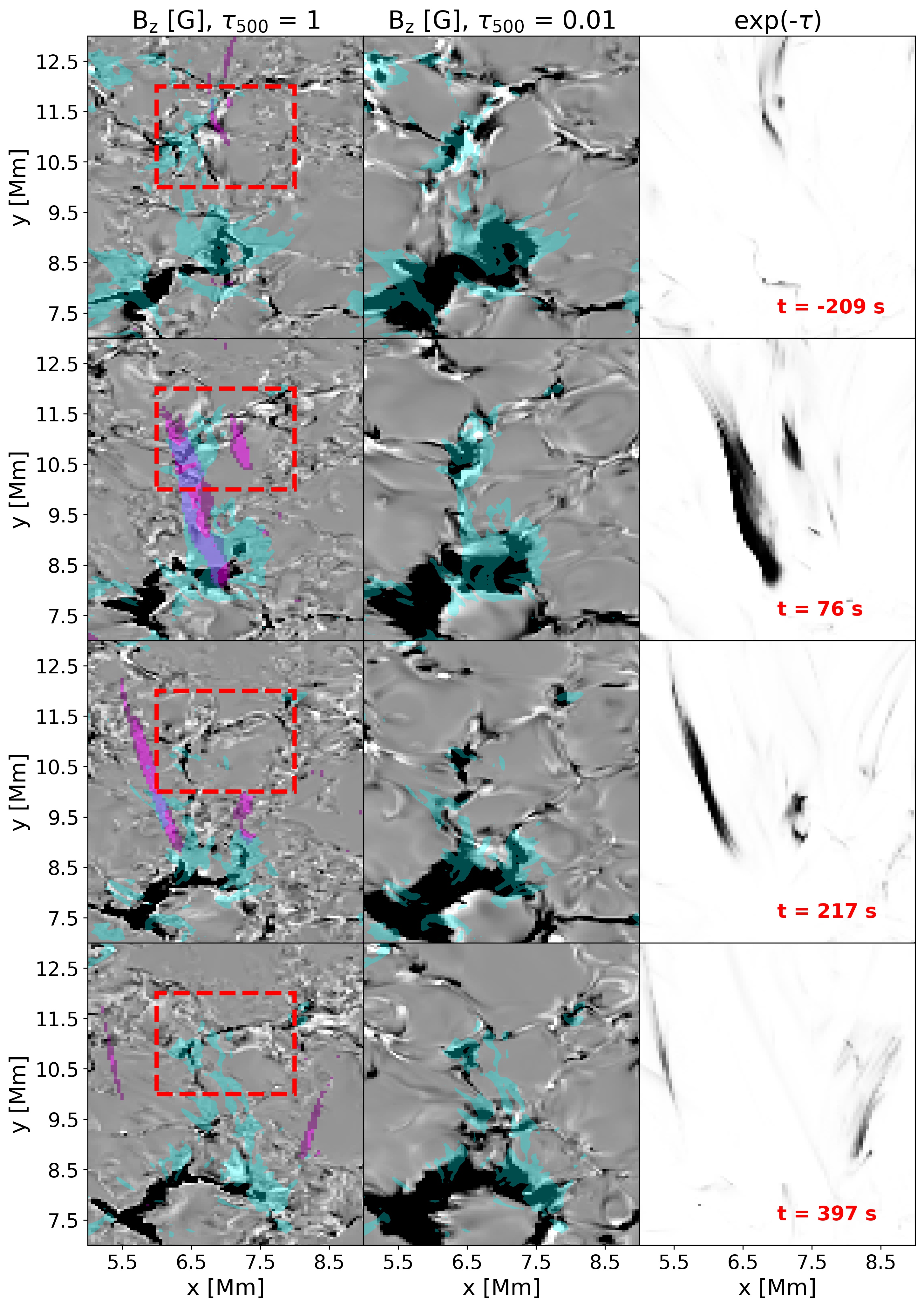}
   \caption{Zoom-in on the evolution of the line-of-sight magnetic field at various $\tau$ levels in the region of the studied feature. Left: Evolution of the vertical component of the magnetic field, \textit{$B_\mathrm{z}$} at the photosphere ($\tau_{500}$ = 1). The feature from the proxy is overplotted in purple. The dashed red box indicates the region of flux cancellation. Centre: Evolution of \textit{$B_\mathrm{z}$} in the middle photopshere ($\tau_{500}$ = 0.01). The grey scale for the magnetic field (\textit{$B_\mathrm{z}$}) ranges from $-500\;$G to $500\;$G at the $\tau_{500}$ = 1 surface and from $-100\;$G to $100\;$G at the $\tau_{500}$ = 0.01 surface. 
   The regions of enhanced current density (75\% above the mean) are plotted with the cyan contours in both the left and central panels. Right: Zoom-in on the evolution of the feature in the proxy as viewed on disc. A movie spanning the lifetime of the feature is available \href{https://owncloud.gwdg.de/index.php/s/cw2KeAC59h7W2DF}{online}.}
              \label{Bztau_zoom}%
\end{figure}

\begin{figure*}
   \centering
   \includegraphics[width=1\textwidth]{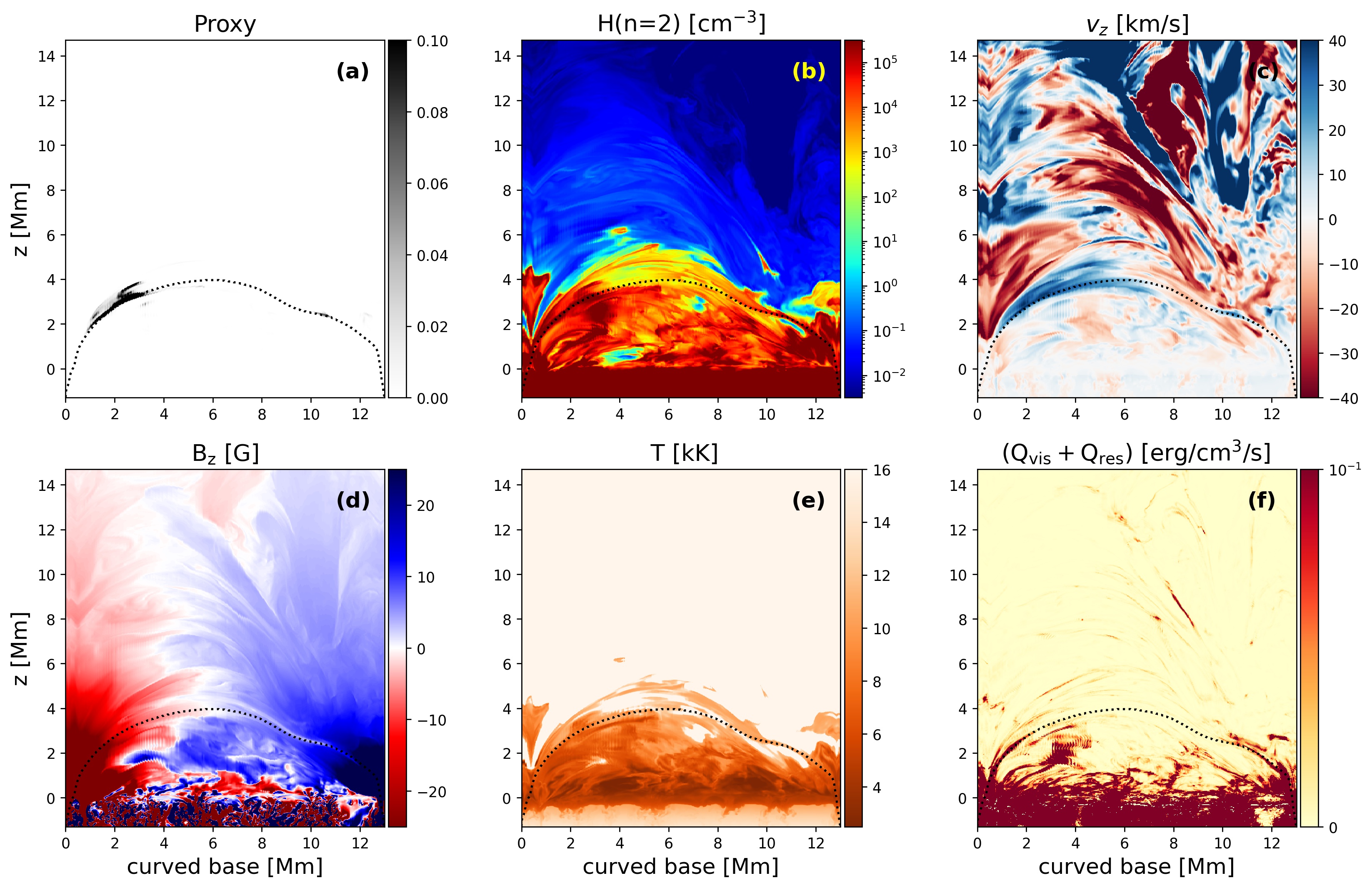}
   \caption{Vertical plane passing through the field line containing the feature. The field line is shown by the dotted black line. A vertical plane with a curved base, containing the field line is shown for various quantities. The base of this plane on the $\tau_{500}=1$ surface is indicated by the dashed green curve in Fig.~\ref{bz-topo-before}. Panel (a) shows the proxy (normalised $\mathrm{\kappa}$) in the vertical plane. Panel (b) is the hydrogen number density at the n = 2 atomic level. Panels (c) and (d) are the vertical velocity and magnetic field components. Panel (e) is the temperature. Panel (f) shows the viscous and resistive heating rate. A time-series for the evolution of this feature and its MHD properties and heating in this plane is available \href{https://owncloud.gwdg.de/index.php/s/Yc4N1yvbU46iGBd}{online}.}
              \label{curved_slice}%
\end{figure*}

\begin{figure*}
   \centering
   \includegraphics[width=1\textwidth]{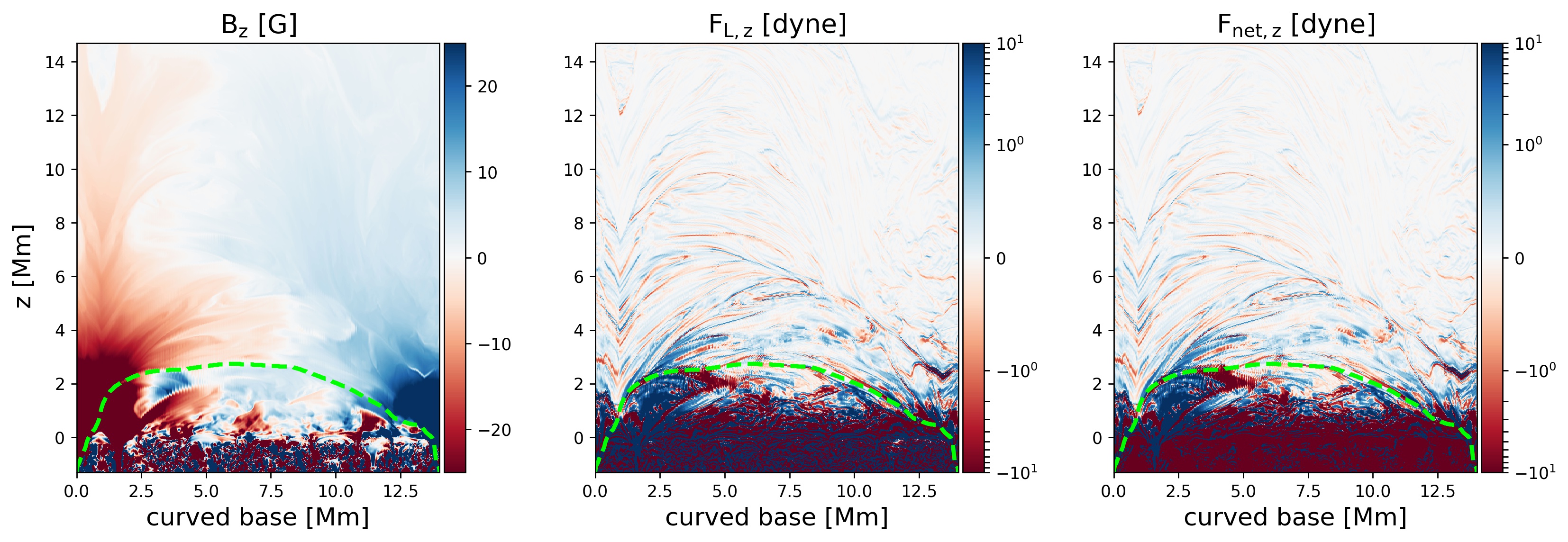}
   \caption{Snapshot showing the vertical field and forces after flux emergence with a highly kinked field. The vertical component of the magnetic field ($\mathrm{B_z}$) and the Lorentz ($\mathrm{F_{L,z}}$), and net forces ($\mathrm{F_{net,z}}$) is shown. The traced field line containing the feature is shown by the dashed green line. The evolution of these forces and fields is available \href{https://owncloud.gwdg.de/index.php/s/fg8IdQcCLoXG3rD}{online}.}
              \label{Lorentz-force}%
\end{figure*}

\begin{figure*}
   \centering
   \includegraphics[width=1\textwidth]{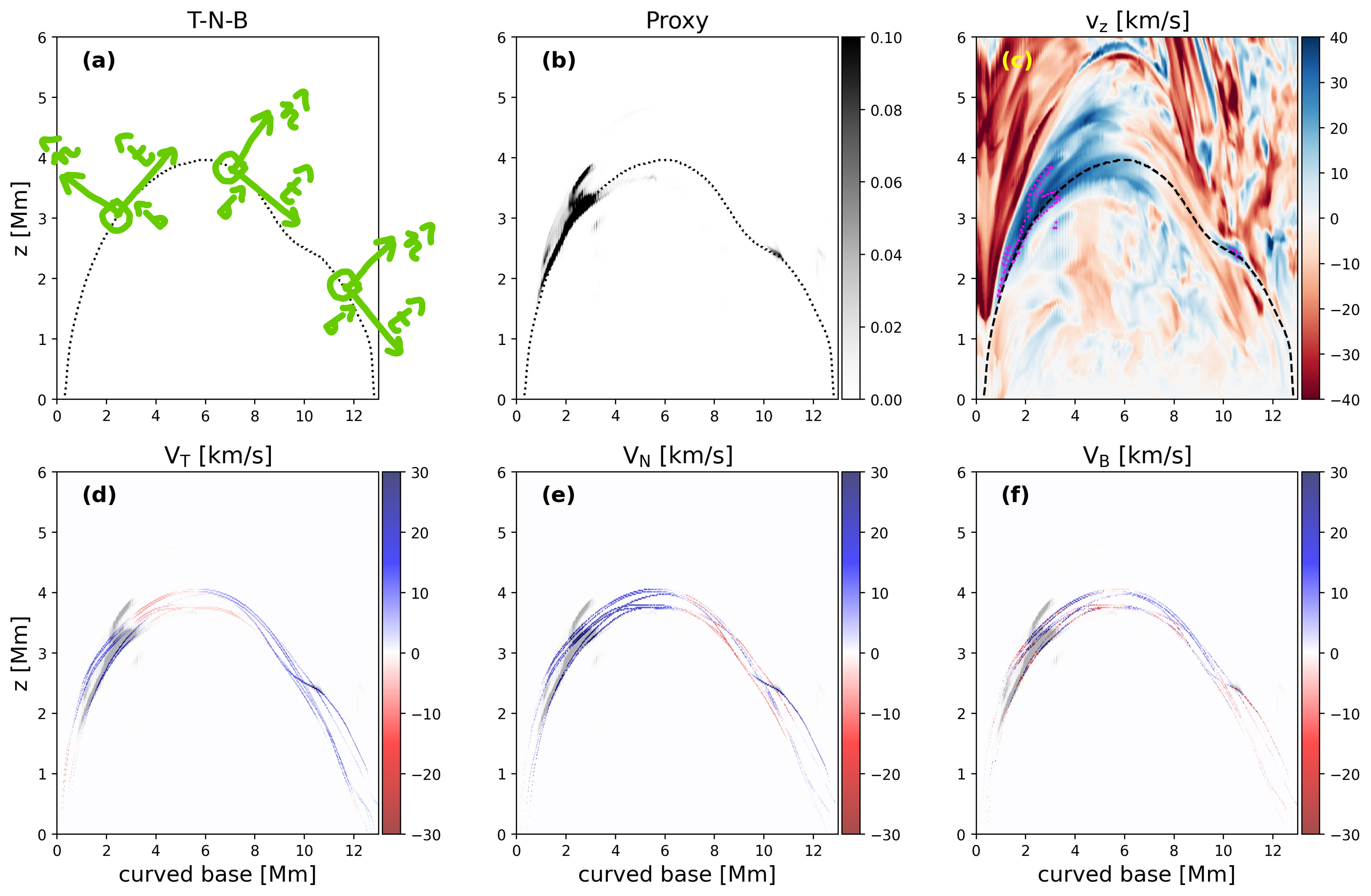}
   \caption{The velocity components driving the feature and its visibility. Panel (a) shows the direction of the tangent (T), normal (N) and binormal (B) along the field line for this snapshot. Panel (b) shows the feature on the field line viewed in the proxy. Panel (c) shows the line-of-sight (vertical) velocity and outlines the feature with the magenta contour. Panels (d)-(f) show the flow components along a bunch of field lines traced through the feature.}
              \label{velocity_cartoon}%
\end{figure*}

\begin{figure*}
   \centering
   \includegraphics[width=1\textwidth]{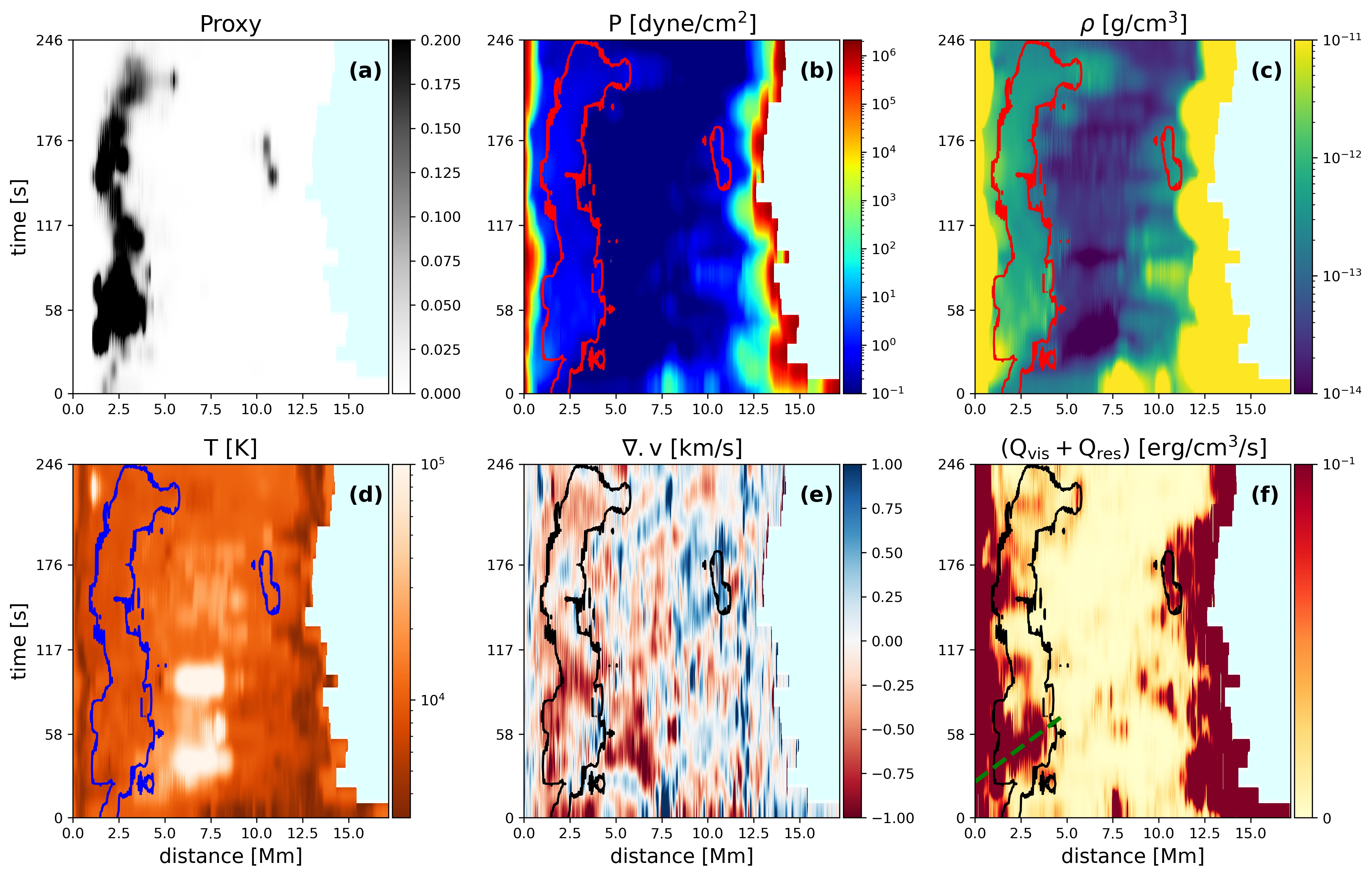}
   \caption{Time-distance plots of various MHD quantities along the selected field line (dashed green line in Fig.~\ref{Lorentz-force}). Panel (a) shows the evolution of the feature in the proxy. Panels (b), (c), and (d) show the gas pressure, density, and temperature, respectively. Panel (e) shows the divergence of flows, highlighting regions of compression (red)/expansion (blue). Panel (f) shows the viscous and resistive heating rate where the dashed green line shows the Alfv\'en speed in the region of the feature at $\sim$ 100 km/s. Contour lines highlight the location of the feature in panels (b) - (f). The length of the traced field lines change with time within the domain, so the data is padded with zeros on the right. This is shown with the light cyan colour.} 
              \label{Time-distance(fundamentals)}%
\end{figure*}

\begin{figure*}
   \centering
   \includegraphics[width=1\textwidth]{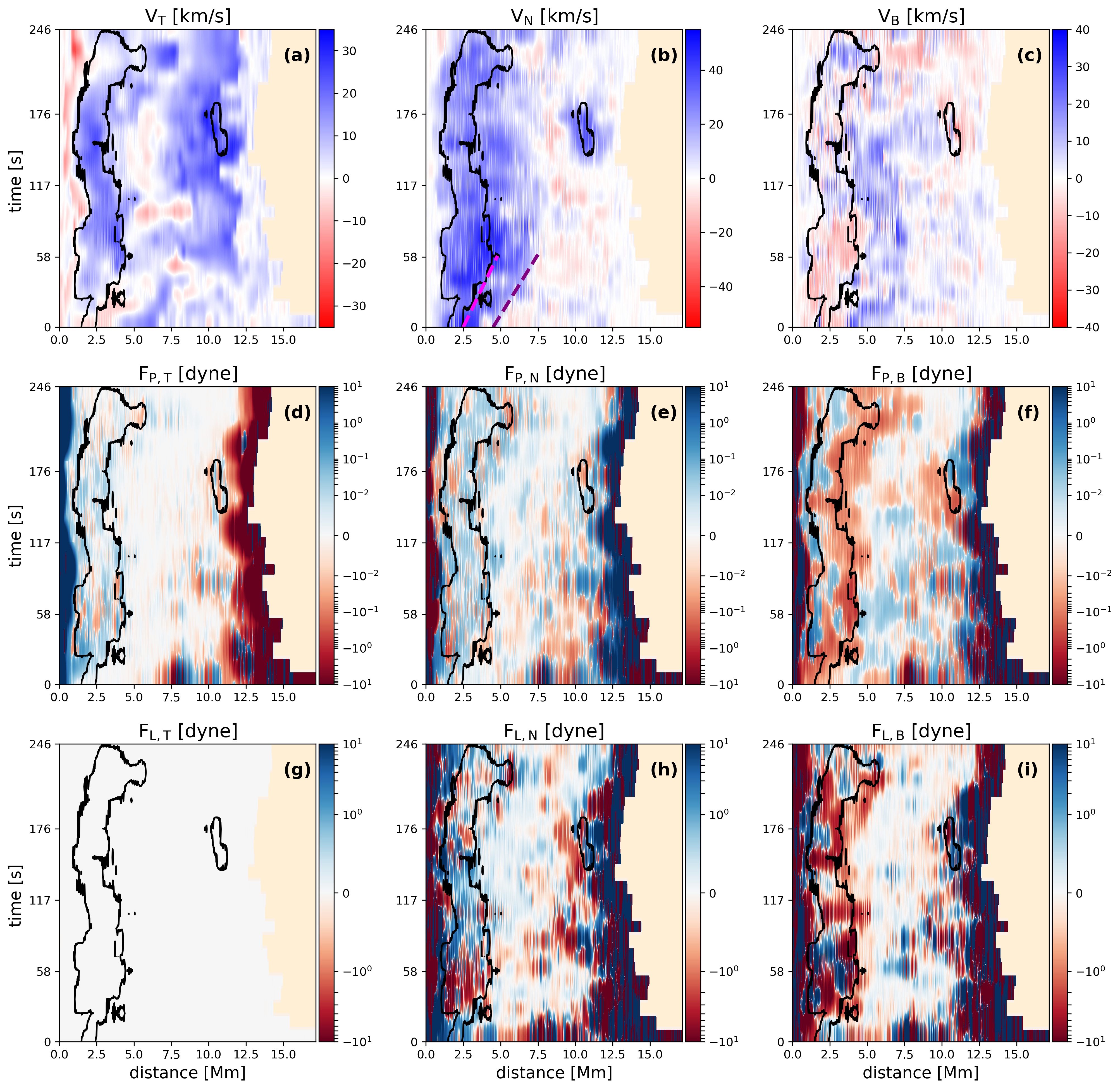}
   \caption{Time-distance plots for the projection of flows and forces in the tangential (T), normal (N) and binormal (B) directions, respectively. Panels (a)-(c): flows projected in the T, N, B directions in the frame of reference of the selected field line. Panels (d)-(f) are the same for the net pressure gradient forces. Panels (g)-(i) show the Lorentz forces projected in the T, N, B directions. The dashed purple line in panel (b) marks the apparent velocity of propagation along the field line which is 50\,km/s. The apparent velocity of the feature itself marked by the dashed magenta line is 38\,km/s. Black contours highlight the location of the feature in all the panels. The padding on the right is indicated by the light orange colour.} 
              \label{Time-distance(flowsforces)}%
\end{figure*}


\section{Analysing a blueshifted feature}\label{feature-of-interest}

 We now present the analysis of an example feature identified in the blue wing of our proxy in the same snapshot as used for the previous analysis. The feature is visible at a Doppler shift of 37\,km/s. It appears as a straight feature at $\mu$ = 1 with a maximum length of 3.4\,Mm and displays a very similar appearance in the image generated by MULTI3D at the same Doppler shift (see panels a and b of Fig. \ref{37F1}). We take a vertical slice (panel d in Fig. \ref{37F1}) through the feature to better gauge how the structure looks in 3D. Although it appears as a straight feature in the on-disc line-of-sight integrated maps, it has a loop-like morphology when viewed in the vertical cuts. Furthermore, the formation height obtained from the H$\alpha$ MULTI3D synthesis (panel c in Fig. \ref{37F1}) shows that the feature forms at about 2-4\,Mm above the average solar surface. This is consistent with what we see in the vertical cut of the proxy feature (panel d in Fig. \ref{37F1}). The proxy is used as the normalised absorption coefficient ($\kappa$) for the vertical slices, which shows the density structure with height. In panel (e) we find that the feature (overplotted in black) lies in an upflow region, which explains why it appears in the blue wing. In panel (f) we see that it appears to be rooted at a negative magnetic field patch from the vertical component of the magnetic field.    

The movie associated with Fig. \ref{37F1} shows that the feature forms in a region outside the chosen vertical slice. The feature forms close to time = 0\,s. We use t = 0 as the time for feature formation henceforth. As the feature evolves, it moves into this vertical plane and exits it again before disappearing. The lifetime of the feature is approximately 4 mins. While the feature is still in the plane, it is observed to expand outwards. It can be seen that any feature detected by the proxy at this Doppler velocity always lies in upflow regions and has cool, chromsospheric temperatures. Furthermore, the hydrogen (n = 2) population densities are locally enhanced in the region of these features. 

A close inspection of the movie (in Fig. \ref{37F1}) reveals that as the feature enters the vertical slice, a positive polarity intrusion is visible in the same region, in a predominantly negative polarity field. This hints at a possibility of magnetic reconnection. We explore this further in the following section \ref{driving_mechanism}. 

\subsection{Driving mechanism}\label{driving_mechanism}

We traced magnetic field lines close to the site of feature formation to better understand possible magnetic reconnection through the rearrangement of fields. The 3D rendering of magnetic field lines reveals an overarching bundle of negative polarity field lines that overlie the site of flux cancellation (see Fig. \ref{bz-topo-before}). Another bundle of field lines with opposite polarity is seen to bend and cross these overarching field lines from below. This is seen as the opposite polarity intrusion in the vertical cuts shown in Fig. \ref{37F1}. This secondary bundle of field lines connects closer to the main footpoint of the feature and are seen to undergo rearrangement as flux emergence and cancellation occurs there. This is highlighted by the red circle. The rearrangement of magnetic field lines at this site is visible in the movie associated with Fig. \ref{bz-topo-before}.

We further looked at the evolution of the line-of-sight (vertical) component of the magnetic field at different optical depths. In Fig. \ref{Bztau_zoom} we show the evolution at two $\tau_{500}$ surfaces zooming into the feature. At $\tau_{500} = 1$, the evolution of photospheric fields show signatures of flux emergence and consequent cancellation. The dashed red box in the leftmost panel in Fig. \ref{Bztau_zoom} highlights the site where magnetic reconnection is likely to occur. This is the same region as marked by the red circle in Fig. \ref{bz-topo-before}. It is clear that the magnetic patch outlined by the dashed red box begins with a predominant positive polarity. A small negative polarity patch emerges close to it and cancels the pre-existing positive polarity magnetic field. Eventually, the positive polarity is completely cancelled and negative polarity fields remain in that region. This is also visible in the middle photosphere ($\tau_{500}$ = 0.01). This indicates that flux cancellation and reconnection also extend into the low chromosphere. Additionally, regions of enhanced current density |\textit{J}|, integrated along the line of sight, are highlighted through cyan shading. The threshold is set to 1.75  times the mean value. This analysis further bolsters the hypothesis that flux cancellation and consequent magnetic reconnection, in the photosphere and chromosphere, drives the formation small-scale features \citep{Shibata_2007, Samanta_2019, Duan_2023}. 

The movie associated with Fig. \ref{Bztau_zoom} also shows some physical characteristics of the feature. As mentioned before it has a maximum length of 3.4\,Mm and lifetime of 246\,s. It also exhibits a lateral displacement (leftward as viewed on-disc) of up to 400\,km during its lifetime. This is consistent with observations of RBEs and RREs \citep{ Sekse_2013, Kuridze_2015, Bose_2021}. 

\subsection{Tracing field lines containing the feature}\label{trace-field}

To better follow the evolution of the feature, we trace one field line threading the feature in each snapshot. We emphasise that the field line is not being traced through time and it is not necessarily the same field line in different snapshots. Rather, we follow the 3D density feature in the proxy for each snapshot and trace the field line based on its location. The seed point in each snapshot for tracing the field line is chosen based on the proxy feature, such that it lies approximately at the centroid of said feature. The integration is then done bi-directionally to trace the field line. A trilinear interpolation scheme is implemented with the RK45 integration. Fig. \ref{curved_slice} shows the feature in the proxy and various MHD quantities in the plane containing the traced field line. This plane has a curved base (dashed green line in Fig.~\ref{bz-topo-before}) based on the grid points where the field line is traced. In our analysis, we treat the negative polarity footpoint as the starting point (L = 0). The other footpoint shifts and changes length from one snapshot to the next.

\subsection{Heating and MHD properties}\label{heat_mhd}

We aim to track the evolution of the feature over time, focusing on its MHD properties. As is outlined in Sect. \ref{trace-field}, the feature is identified using the proxy in vertical cuts, and its movement is traced from one frame to the next. To study the feature in detail, we followed the field line it resides on and analysed its properties through a vertical plane intersecting this line, as is shown in Fig. \ref{curved_slice}.

The feature, as seen in the proxy, appears elongated and aligned with the magnetic field line. In this region, we observe an increase in the hydrogen populations at the n=2 energy level, which is expected for enhanced absorption in H$\alpha$. The temperature in this area is characteristic of chromospheric temperatures (8\,kK - 12\,kK). Additionally, the feature is located in an upflow region with a vertical velocity of 30 - 40\,km/s.

In panel (d) of Fig. \ref{curved_slice}, the footpoint of the field line carrying the feature is rooted at a negative polarity patch. The footpoints of the overarching field lines discussed in Sect. \ref{driving_mechanism} and Fig. \ref{bz-topo-before} lie in the same magnetic patch. However, the positive flux emerges much closer --- at a distance of about 2 Mm from the negative polarity footpoint. This is the same region highlighted by the red circle and red box in Fig. \ref{bz-topo-before} and Fig. \ref{Bztau_zoom}, respectively. 

In the bottom right panel of Fig. \ref{curved_slice} we plot the viscous and resistive heating rate. Heating is visible at the footpoints and partly in the region of the feature for this snapshot. When we track the heating through time in the linked movie, we can see strong heating at the onset of the feature, associated with flux emergence. There is also intermittent heating at the footpoints as the feature evolves. While the heating appears to show correlation with the feature's development, this is not as apparent in the temperature distribution, which remains relatively homogeneous.

\subsection{Lorentz forces associated with flux emergence}\label{flux-lorentz}

We explore the role of the Lorentz forces associated with the flux emergence in the region of our interest. A detailed description of the Lorentz force treatment in MURaM can be found in \citep{Rempel_2017}. Figure~\ref{Lorentz-force} and the associated animation show the evolution of the vertical component of the magnetic field in the plane of the traced field line. We also show the evolution of the Lorentz forces and the net force from the momentum equation in the vertical (z) direction. The emergence of positive polarity fields is visible close to the negative polarity footpoint of the feature. As the field emerges, it drives reconnection and the traced field line (dashed green line) is pushed upwards (see the movie associated with Fig. \ref{Lorentz-force}). Strong positive Lorentz forces act close to the L = 0 footpoint of the feature. As the emerged flux meets the traced field line, negative Lorentz forces temporarily hold down the field line from expanding. At t $\sim$ 90\,s, the field line is suddenly seen to expand significantly and the Lorentz force then becomes positive. The net force is dominated by the Lorentz force component ($\mathrm{F_L}$) above the solar surface. 
 
\subsection{Frame of reference of the field line}

To analyse the temporal behaviour of the feature, we move to the reference frame of the traced 3D magnetic field line. This allows us to examine how the feature evolves both along the field line’s direction and perpendicular to it.

\subsubsection{Geometry for our analysis}\label{setup}

We define three mutually orthogonal directions: tangents (T), normals (N) and binormals (B). Such a decomposition has been done in previous works to study flows and forces along and perpendicular to the field lines (e.g. \citealp[]{Leenaarts_2015}). The tangents follow the magnetic field. The direction was chosen from the negative footpoint (L = 0) directed towards the positive footpoint. This was done to ensure that the starting point for our analysis is the negative polarity patch where the feature is rooted. The orthogonal set of basis vectors for this frame of reference was constructed along each grid point, s = (x,y,z) of the traced field line. The set of tangent vectors is defined as

\begin{align}
    & & \mathrm{\mathbf{\hat{t}}(s) = -\mathbf{\mathbf{\frac{B}{|B|}}}(s)}.
\end{align}

We obtained the set of binormal vectors, $\mathrm{\mathbf{\hat{b}}(s)}$, with the following relation, keeping in mind the periodic boundary condition in the horizontal directions:

\begin{align}\label{binormal}
    & & \mathrm{\mathbf{\hat{b}}(s) = \pm \ \mathbf{\hat{t}}(s) \times \frac{d\mathbf{t}}{ds}}.
\end{align}

\noindent The normals, $\mathrm{\mathbf{\hat{n}}(s)}$, are obtained with

\begin{align}\label{vector-relation}
    & & \mathrm{\mathbf{\hat{n}}(s) = \mathbf{\hat{b}}(s) \times \mathbf{\hat{t}}(s)}.
\end{align}
   
We have a closed field line at all times, where consecutive tangent vectors are not parallel. This allows this basis construction to be used effectively to move to the frame of reference of the traced field line. 

We chose the sign in Eq. (\ref{binormal}) such that $\mathrm{\mathbf{\hat{b}.\hat{x}} > 0}$. This results in $\mathrm{\mathbf{\hat{n}.\hat{z}} > 0}$. It is to be noted that the field line is always roughly oriented along $\mathrm{\mathbf{\hat{y}}}$. Figure \ref{velocity_cartoon} shows a cartoon of the directions T-N-B along a traced field line in panel (a), with the feature in the proxy on the same field line shown in panel (b).

\subsubsection{Time-distance analysis}

Using the set-up to analyse quantities within the field line’s reference frame, we constructed time-distance plots spanning the feature’s entire lifetime. Figure \ref{Time-distance(fundamentals)} illustrates the feature’s evolution in the proxy, along with various MHD properties along the field line. It is important to note that the length of the field line changes over time. This is also indicated in Figs.~\ref{Time-distance(fundamentals)} and \ref{Time-distance(flowsforces)}.

The feature in our proxy exhibits an oscillatory behaviour over time, as is illustrated in panel (a) of Fig. \ref{Time-distance(fundamentals)} which shows a clear S-shaped evolution. This indicates that the feature moves back and forth along the field line during its lifetime, a motion that is also visible in the close-up movie linked to Fig. \ref{Bztau_zoom}. The period of these oscillations is about 3 minutes. While turbulent motions near the surface displace the footpoint of this feature, the feature’s movement along the field line is notably larger and thus significant. Panels (b) and (c) in Fig. \ref{Time-distance(fundamentals)} reveal increased gas pressure and density within the region of the feature. However, panel (d) shows no corresponding increase in temperature. In panel (e), we see the flow field’s divergence, with the black-contoured feature region distinctly dominated by compression.

We also follow the heating of the feature in panel (f) of Fig.~ \ref{Time-distance(fundamentals)}. This is the viscous and resistive heating rate along the field line. We find clear signatures of co-spatial and co-temporal heating at the birth of the feature. There is a propagating heating front which is marked by the dashed green line in panel (f) at a speed of 100 km/s. The Alfv\'en speed in that region also has a similar value.

We analysed the net forces and flows projected in the tangent (T), normal (N) and binormal (B) directions as described in Sect.~ \ref{setup}. Figure \ref{Time-distance(flowsforces)} presents the component-wise analysis using vector fields --- the flow field (panels a-c), pressure gradient force field (panels d-f), and Lorentz force field (panels g-i).

The tangential velocity ($\mathrm{V_T}$; panel a, Fig. \ref{Time-distance(flowsforces)}) is predominantly positive within the feature (outlined by black contours). This means that the flow consistently moves away from the footpoint. The pressure gradient force along the tangential direction (panel d) is also mostly positive, indicating that pressure differences help drive the flow along the field line.

The normal velocity component ($\mathrm{V_N}$; panel b) is strongly positive in the region of the feature, suggesting an upward and outward flow. This again is driven by the pressure gradient force (panel e), which acts to push the field line. The propagation front (marked by the dashed purple line in panel b) is about 50 km/s, which is faster than the apparent velocity of the feature at 38 km/s (dashed magenta line).

The transverse or binormal component of the flow field $\mathrm{V_B}$ exhibits a mix of positive and negative values, indicating a more complex lateral motion. The pressure gradient force (panel f) shows negative values at the feature’s tip, suggesting that this component could contribute to leftward lateral movement (as seen in the movie associated with Fig.~\ref{Bztau_zoom}).

The treatment of the Lorentz force is similar to that of the curved vertical slices described in Sect. \ref{flux-lorentz}. The forces acting normal to the field line (panel h, Fig. \ref{Time-distance(flowsforces)}) show that strong positive forces precede the prominent appearance of the feature in the proxy. As the feature forms prominently, we find some negative forces associated with the emerged flux interacting with the field line. At about t $\sim$ 90\,s, the force becomes positive again, aiding in the field line expansion (also see Sect. \ref{flux-lorentz}). The binormal component (panel i, Fig. \ref{Time-distance(flowsforces)}) of the Lorentz force shows some domination of negative forces.

\section{Discussion and conclusions}\label{discussion}

We have developed a computationally efficient proxy for identifying Doppler-shifted H$\alpha$ fine structure in MURaM-ChE simulations. 
The features identified by the proxy were validated with MULTI3D synthesis. By approximating key aspects of the NLTE RT problem, the proxy bypasses the need for full 3D NLTE synthesis. This reduces the computational costs from several hundred thousand CPU hours per snapshot to a few CPU seconds. The proxy allows us to capture the morphology and position of chromospheric features appearing in the wings of the spectral line, effectively covering Doppler shifts from approximately 77\,km/s down to 17\,km/s in either wing. The proxy is physically motivated in terms of capturing the interaction of the plasma with the radiation to identify features in the solar atmosphere visible in  the H$\alpha$ wings.  Although the proxy focuses on features in the wings of H$\alpha$ and does not model the line core, it has shown reliable accuracy in identifying TPs and TNs within its scope. Because the proxy is very fast, it allows us to identify many such features, so that the statistical properties of these features can be studied, which will be the topic of a follow-up paper.

The proxy assumes a Gaussian profile for the spectral dependence of the escape probability. To accurately model the features in the far wings where pressure broadening becomes important, a Voigt profile is potentially more fitting for the construction of the proxy. The average escape probability profile from the MULTI3D synthesis matches the Gaussian profile closely. Thus the approximation of using a Doppler profile to model the proxy works well. We also found that the maximum difference of the fits to the average exp(-$\tau$) profile of the synthetic H$\alpha$ line (Fig. \ref{Fitting}) with a Voigt profile and a Gaussian profile is less than 10\% of the maximum.

The RBEs and RREs identified using the proxy are similar in morphology and evolution to those observed. These are fine needle-like projections seen on the solar disc and appear to be jets. We also identify some fibrillar structures which are broader and more smudged. All these features appear to be rooted close to the network patch in our simulations, similar to observations in H$\alpha$ and Ca II H \& K.

The example feature that was investigated is an RBE at a Doppler shift of 37\,km/s. We infer the morphology of the feature from two different vantage points --- (1) the on-disc or top view and (2) the vertical cuts. In essence, we use the absorption coefficient directly (in vertical cuts) or to construct the escape probability (in the top view) to understand the feature in 3D. The on-disc map shows a nearly straight feature evolving on the solar disc. This gives an impression that it is a straight, jet-like feature. The proxy in the vertical cuts shows that we are actually looking near the top of a closed loop structure. Given that the feature is pushed along the field line, we observe a jet-like behaviour. The loop itself also moves upwards and expands carrying the feature along with it. However, it is different from a traditional type-II spicule, where we would expect a straight feature to shoot up in a linear trajectory. Whether the feature does appear as a type II spicule from a different viewing angle is a matter of future investigation. These aspects further highlight that studying the 3D morphology of these features is important for a comprehensive understanding of their behaviour. It also highlights the need to determine if there are fundamental differences between spicules and RBEs/RREs.

 The feature has a maximum length of $3.4~\mathrm{Mm}$ during its lifetime spanning 4 minutes. It also exhibits lateral displacement ($\sim~400~\mathrm{km}$). The height of formation for this feature is $2-4~\mathrm{Mm}$, corresponding to heights within the chromosphere. The proxy identifies the correct height of formation, consistent with the result from MULTI3D. Our time-distance analysis shows that in the region of the feature the temperature ranges from 8\,kK to 10\,kK on average, corresponding to the chromosphere and low transition region. The density ranges from $10^{-13}~\mathrm{g/cm^3}$ to $10^{-12}~\mathrm{g/cm^3}$. The pressure in the region of the feature is approximately $0.1~\mathrm{dyne/cm^2}$ to $1~\mathrm{dyne/cm^2}$. While the density, pressure and hydrogen (n = 2) densities are enhanced in the region of the feature, the temperature does not show this local enhancement. This concurs to the fact that the H$\alpha$ opacity is only weakly sensitive to the temperature \citep{Leenaarts_2012}.

The formation of the RBE is driven by flux emergence and magnetic reconnection. The flux emergence process pushes the plasma upwards. Reconnection follows and a convergence of flows in the region of the feature is clearly visible (panel f in Fig.~\ref{Time-distance(fundamentals)}). The signature of a bi-directional flow is visible at the site of reconnection (boundary of the feature in panel a, Fig. \ref{Time-distance(flowsforces)}). The primary force behind the field-aligned flows acting on the feature is the pressure gradient. The reconnection outflow (panel a, Fig.~\ref{Time-distance(flowsforces)}) hits previously slow-moving gas and leads to compression, resulting in the local enhancement of the plasma density and pressure. Pressure gradient forces build up and this drives a flow along the field line carrying the feature. This makes our example feature behave like a jet. The pressure gradient forces also lead to strong flows acting on the field line itself pushing it upwards, causing it to expand. We also find strong Lorentz forces associated with the flux emergence process. The transverse component of the forces may be responsible for the lateral motion ($\sim~400~\mathrm{km}$) seen in the feature. 

The time-distance analysis of the flow component perpendicular (outwards) to the field line reveals actual mass flows at speeds of $37-39~\mathrm{km/s}$ associated with the feature. In addition, we find a propagation front at $50~\mathrm{km/s}$. This front may correspond to faster features appearing in the same region as our target feature. It could also indicate flows which are not associated with mass flows, but rather wave modes propagating at sound speeds. 

It is to be noted that we have identified features based on the line-of-sight, that is, vertical component of the velocity. Hence we detect the feature only in upflow regions of $\sim$ 37 km/s (see Fig. \ref{velocity_cartoon}). However, the flow speeds in Fig.~\ref{velocity_cartoon} along the field line ($\mathrm{v_T}$) are much slower ($\sim$ 7 km/s). Although there are strong upflows, the region occupied by the upflows and the plasma  does not move significantly. Hence the jet appears relatively stationary. 

We can see that bidirectional flows occur throughout the feature's lifetime (panel a, Fig. \ref{Time-distance(flowsforces)}), and the field remains kinked (movie associated with Fig. \ref{Lorentz-force}). Thus, the reconnection appears to be powering the feature for the most part of its lifetime. 

An intriguing propagation front that we observe in our analysis comes from the propagation of the viscous and resistive heating that travels at Alfv\'enic speeds of $\sim$ 100 km/s. The strong viscous and resistive heating close to the first appearance of the feature exhibits this heating front. The footpoints of the field line carrying the feature are also heated intermittently. While this heating could possibly be unrelated to the feature itself, it shows a good correlation with the flux emergence event driving the feature formation (see the movie linked to Fig.\,\ref{curved_slice}). The heating at the footpoints show no associated feature formation. This could partly be because there are no associated features present, or because we do not identify them at the Doppler velocity used for this study. There have been conjectures on rapidly propagating heating fronts associated with spicules that extend into the corona \citep{de_Pontieu_2011, Bose_2021}. What we observe could be one of these propagating heating fronts responsible for transporting the energy from the chromosphere into the corona, thus being one of the viable mechanisms of coronal heating. However, we do not find a multi-thermal nature associated with the feature unlike in \cite{Bose_2021}, although they study a downflowing feature.

We do not find the heating to manifest itself in the temperature. Radiative losses and ionisation processes may be responsible for this. The energy could also be used to increase the kinetic energy of the plasma, driving mass motions rather than a temperature increase. We surmise the latter to be the case. 

\cite{Rutten_2017} argues that the high opacity of H$\alpha$ fibrils is caused by hot precursors called propagating heating events or PHEs. The fibrils themselves are contrails of such fast-propagating heating events. RBEs also fall in the family of features that PHEs give rise to. Although we do not trace back to the conditions before the formation of the feature, we find the strong heating front only after the flux has emerged and interacts with the field line. This happens co-temporally with the  appearance of the studied feature. We argue that magnetic reconnection events lead to viscous dissipation generating the heating in this feature. The formation of the feature happens simultaneously with --- not following the heating event.

 We also find an oscillatory behaviour: the feature moves back and forth along the magnetic field line carrying it.  The period of these oscillations is about 3 minutes. This is visible in the S-shaped evolution of the feature in panel a, Fig.~\ref{Time-distance(fundamentals)} and the movie associated with Fig.~\ref{Bztau_zoom}. We find a lateral displacement of a few hundred kilometres of the feature in a leftward direction in the on-disc maps. This is consistent with observations of RBEs \citep{Sekse_2012,Sekse_2013}. Transverse displacements of RBEs throughout the entire length of the feature have also been observed \citep{vanderVoort_2009,Kuridze_2015}. In our case, it appears to be the displacement of the entire flux tube at all heights as speculated by \cite{Kuridze_2015}. In a numerical study of fibrils by \cite{Druett_2022}, they argue that the gas pressure gradient plays a dominant role in such lateral displacements. Our findings are in sync with this. We find that it is a combination of both the gas pressure gradient and the Lorentz forces acting on the flux tube that moves it in a general leftward (-x) direction in the on-disc view.

 We successfully observed many features similar to RBEs and RREs without the use of ambipolar diffusion in the simulations. \cite{Bose_2021} found ambipolar diffusion \citep{Sykora_2017} to play a major role in heating with their Bifrost \citep{Gudiksen_2011} simulations. It could be that the exclusion of ambipolar diffusion led to an insignificant impact on our temperatures. However, the formation of jet-like features in simulations done with the MURaM-ChE code, with properties similar to observed RBEs, does not necessitate ambipolar diffusion to be incorporated.

 The RBEs and RREs analysed by \cite{Danilovic_2023} exhibit several observed properties -- such as spatial distribution, lateral motions, lengths, and lifetimes -- that closely align with our findings. Additionally, their reported line-of-sight velocities ($30-40~\mathrm{km/s}$) and formation heights ($2-4~\mathrm{Mm}$) are consistent with the values we observe.

The proxy can be developed further to work at the solar limb to study spicules. This will be presented in a future work. Devising a proxy for the line core of H$\alpha$ to study dynamic fibrils is not straightforward. As was elucidated by \cite{Leenaarts_2012}, the line core has nuances which are difficult to model. Modelling the entire H$\alpha$ spectral line at a low computational cost remains a challenge.

\begin{acknowledgements}
       We thank the anonymous referee for valuable comments that helped improve the paper. The authors also thank Yajie Chen for useful discussions. This work was carried out in the framework of the International Max Planck Research School (IMPRS) at the Technical University of Braunschweig. We are grateful for the computational resources provided by the Cobra and Raven supercomputer systems of the Max Planck Computing and Data Facility (MPCDF) in Garching, Germany. This research has received financial support from the European Union’s Horizon 2020 research and innovation program under grant agreement No. 824135 (SOLARNET) and through the European Research Council (ERC) No. 101097844 (WINSUN). This work was supported by the Deutsches Zentrum f{\"u}r Luft und Raumfahrt (DLR; German Aerospace Center) by grant DLR-FKZ 50OU2201. This project has received funding from the Swedish Research Council (2021-05613) and the Swedish National Space Agency (2021-00116). We acknowledge resources provided by the National Academic Infrastructure for Supercomputing in Sweden (projects NAISS 2023/1-15 and NAISS 2024/1-14) at the PDC Centre for High Performance Computing (PDC-HPC) at the Royal Institute of Technology.
\end{acknowledgements}
%
\bibliographystyle{aa} 
\bibliography{bibfile} 

\begin{thebibliography}{46}
\expandafter\ifx\csname natexlab\endcsname\relax\def\natexlab#1{#1}\fi

\bibitem[{Beckers(1968)}]{Beckers_1968}
Beckers, J.~M. 1968, Solar Physics, 3, 367

\bibitem[{{Beckers}(1972)}]{Beckers_1972}
{Beckers}, J.~M. 1972, \araa, 10, 73

\bibitem[{Bose {et~al.}(2021)Bose, Joshi, Henriques, \& van~der
  Voort}]{Bose_2021}
Bose, S., Joshi, J., Henriques, V. M.~J., \& van~der Voort, L.~R. 2021,
  Astronomy \& Astrophysics, 647, A147

\bibitem[{{Carlsson} {et~al.}(2019){Carlsson}, {De Pontieu}, \&
  {Hansteen}}]{Carlsson_2019}
{Carlsson}, M., {De Pontieu}, B., \& {Hansteen}, V.~H. 2019, \araa, 57, 189

\bibitem[{Chaurasiya {et~al.}(2024)Chaurasiya, Bayanna, Louis, Pereira, \&
  Mathew}]{Chaurasiya_2024}
Chaurasiya, R., Bayanna, A.~R., Louis, R.~E., Pereira, T. M.~D., \& Mathew,
  S.~K. 2024, The Astrophysical Journal, 970, 179

\bibitem[{Danilovic {et~al.}(2023)Danilovic, Bjørgen, Leenaarts, \&
  Rempel}]{Danilovic_2023}
Danilovic, S., Bjørgen, J.~P., Leenaarts, J., \& Rempel, M. 2023, Astronomy \&
  Astrophysics, 670, A50

\bibitem[{{De Pontieu} {et~al.}(2004){De Pontieu}, {Erd{\'e}lyi}, \&
  {James}}]{de_Pontieu_2004}
{De Pontieu}, B., {Erd{\'e}lyi}, R., \& {James}, S.~P. 2004, \nat, 430, 536

\bibitem[{De~Pontieu {et~al.}(2011)De~Pontieu, McIntosh, Carlsson, Hansteen,
  Tarbell, Boerner, Martinez-Sykora, Schrijver, \& Title}]{de_Pontieu_2011}
De~Pontieu, B., McIntosh, S., Carlsson, M., {et~al.} 2011, Science, 331, 55

\bibitem[{De~Pontieu {et~al.}(2007)De~Pontieu, McIntosh, Hansteen, Carlsson,
  Schrijver, Tarbell, Title, Shine, Suematsu, Tsuneta, Katsukawa, Ichimoto,
  Shimizu, \& Nagata}]{de_Pontieu_2007b}
De~Pontieu, B., McIntosh, S., Hansteen, V.~H., {et~al.} 2007, Publications of
  the Astronomical Society of Japan, 59, S655

\bibitem[{Druett {et~al.}(2022)Druett, Leenaarts, Carlsson, \&
  Szydlarski}]{Druett_2022}
Druett, M.~K., Leenaarts, J., Carlsson, M., \& Szydlarski, M. 2022, Astronomy
  \& Astrophysics, 665, A6

\bibitem[{Duan {et~al.}(2023)Duan, Shen, Chen, Tang, Zhou, Zhou, \&
  Tan}]{Duan_2023}
Duan, Y., Shen, Y., Chen, H., {et~al.} 2023, The Astrophysical Journal Letters,
  942, L22

\bibitem[{Gudiksen {et~al.}(2011)Gudiksen, Carlsson, Hansteen, Hayek,
  Leenaarts, \& Martínez-Sykora}]{Gudiksen_2011}
Gudiksen, B.~V., Carlsson, M., Hansteen, V.~H., {et~al.} 2011, Astronomy \&
  Astrophysics, 531, A154

\bibitem[{{Hummer} \& {Rybicki}(1982)}]{Hummer_1982}
{Hummer}, D.~G. \& {Rybicki}, G.~B. 1982, \apj, 254, 767

\bibitem[{Krikova {et~al.}(2023)Krikova, Pereira, \& Rouppe van~der
  Voort}]{Krikova_2023}
Krikova, K., Pereira, T. M.~D., \& Rouppe van~der Voort, L. H.~M. 2023,
  Astronomy \&; Astrophysics, 677, A52

\bibitem[{{Kuridze} {et~al.}(2015){Kuridze}, {Henriques}, {Mathioudakis},
  {Erd{\'e}lyi}, {Zaqarashvili}, {Shelyag}, {Keys}, \& {Keenan}}]{Kuridze_2015}
{Kuridze}, D., {Henriques}, V., {Mathioudakis}, M., {et~al.} 2015, \apj, 802,
  26

\bibitem[{Langangen {et~al.}(2008)Langangen, De~Pontieu, Carlsson, Hansteen,
  Cauzzi, \& Reardon}]{Langangen_2008}
Langangen, {\O}., De~Pontieu, B., Carlsson, M., {et~al.} 2008, The
  Astrophysical Journal, 679, L167–L170

\bibitem[{{Leenaarts} \& {Carlsson}(2009)}]{Leenaarts_and_Carlsson}
{Leenaarts}, J. \& {Carlsson}, M. 2009, in Astronomical Society of the Pacific
  Conference Series, Vol. 415, The Second Hinode Science Meeting: Beyond
  Discovery-Toward Understanding, ed. B.~{Lites}, M.~{Cheung}, T.~{Magara},
  J.~{Mariska}, \& K.~{Reeves}, 87

\bibitem[{Leenaarts {et~al.}(2012)Leenaarts, Carlsson, \& Rouppe van~der
  Voort}]{Leenaarts_2012}
Leenaarts, J., Carlsson, M., \& Rouppe van~der Voort, L. 2012, The
  Astrophysical Journal, 749, 136

\bibitem[{{Leenaarts} {et~al.}(2015){Leenaarts}, {Carlsson}, \& {Rouppe van der
  Voort}}]{Leenaarts_2015}
{Leenaarts}, J., {Carlsson}, M., \& {Rouppe van der Voort}, L. 2015, \apj, 802,
  136

\bibitem[{{Mart{\'\i}nez-Sykora} {et~al.}(2017){Mart{\'\i}nez-Sykora}, {De
  Pontieu}, {Hansteen}, {Rouppe van der Voort}, {Carlsson}, \&
  {Pereira}}]{Sykora_2017}
{Mart{\'\i}nez-Sykora}, J., {De Pontieu}, B., {Hansteen}, V.~H., {et~al.} 2017,
  Science, 356, 1269

\bibitem[{Martínez-Sykora {et~al.}(2009)Martínez-Sykora, Hansteen,
  De~Pontieu, \& Carlsson}]{Sykora_2009}
Martínez-Sykora, J., Hansteen, V., De~Pontieu, B., \& Carlsson, M. 2009, The
  Astrophysical Journal, 701, 1569–1581

\bibitem[{Neckel \& Labs(1984)}]{Neckel_1984}
Neckel, H. \& Labs, D. 1984, Solar Physics, 90

\bibitem[{Ondratschek {et~al.}(2024)Ondratschek, Przybylski, Smitha, Cameron,
  Solanki, \& Leenaarts}]{Ondratschek_2024}
Ondratschek, P., Przybylski, D., Smitha, H.~N., {et~al.} 2024, Mg ii h\&k
  spectra of an enhanced network region simulated with the MURaM-ChE code.
  Results using 1.5D synthesis

\bibitem[{Pereira {et~al.}(2012)Pereira, De~Pontieu, \&
  Carlsson}]{Pereira_2012}
Pereira, T. M.~D., De~Pontieu, B., \& Carlsson, M. 2012, The Astrophysical
  Journal, 759, 18

\bibitem[{Pereira {et~al.}(2016)Pereira, Voort, \& Carlsson}]{Pereira_2016}
Pereira, T. M.~D., Voort, L. R. v.~d., \& Carlsson, M. 2016, The Astrophysical
  Journal, 824, 65

\bibitem[{{Pikel'Ner}(1971)}]{Pikelner_1971}
{Pikel'Ner}, S.~B. 1971, Comments on Astrophysics and Space Physics, 3, 33

\bibitem[{Przybylski {et~al.}(2022)Przybylski, Cameron, Solanki, Rempel,
  Leenaarts, Anusha, Witzke, \& Shapiro}]{Przybylski_2022}
Przybylski, D., Cameron, R., Solanki, S.~K., {et~al.} 2022, Astronomy \&
  Astrophysics, 664, A91

\bibitem[{{Rempel}(2017)}]{Rempel_2017}
{Rempel}, M. 2017, \apj, 834, 10

\bibitem[{{Rouppe van der Voort} {et~al.}(2009){Rouppe van der Voort},
  {Leenaarts}, {de Pontieu}, {Carlsson}, \& {Vissers}}]{vanderVoort_2009}
{Rouppe van der Voort}, L., {Leenaarts}, J., {de Pontieu}, B., {Carlsson}, M.,
  \& {Vissers}, G. 2009, \apj, 705, 272

\bibitem[{Rouppe van~der Voort {et~al.}(2009)Rouppe van~der Voort, Leenaarts,
  de~Pontieu, Carlsson, \& Vissers}]{Rouppe_van_der_Voort_2009}
Rouppe van~der Voort, L., Leenaarts, J., de~Pontieu, B., Carlsson, M., \&
  Vissers, G. 2009, The Astrophysical Journal, 705, 272–284

\bibitem[{Rutten(2007)}]{Rutten_2007}
Rutten, R.~J. 2007, Observing the Solar Chromosphere

\bibitem[{{Rutten}(2008)}]{Rutten_2008}
{Rutten}, R.~J. 2008, in Astronomical Society of the Pacific Conference Series,
  Vol. 397, First Results From Hinode, ed. S.~A. {Matthews}, J.~M. {Davis}, \&
  L.~K. {Harra}, 54

\bibitem[{Rutten(2017)}]{Rutten_2017}
Rutten, R.~J. 2017, Astronomy \& Astrophysics, 598, A89

\bibitem[{{Rybicki} \& {Lightman}(1979)}]{Rybicki_1979}
{Rybicki}, G.~B. \& {Lightman}, A.~P. 1979, {Radiative processes in
  astrophysics}

\bibitem[{Samanta {et~al.}(2019)Samanta, Tian, Yurchyshyn, Peter, Cao,
  Sterling, Erdélyi, Ahn, Feng, Utz, Banerjee, \& Chen}]{Samanta_2019}
Samanta, T., Tian, H., Yurchyshyn, V., {et~al.} 2019, Science, 366, 890–894

\bibitem[{Secchi(1871)}]{secchi_1871}
Secchi, A. 1871, Anno XXIV, Sessione III del, 7

\bibitem[{Sekse {et~al.}(2013{\natexlab{a}})Sekse, Rouppe van~der Voort, \&
  De~Pontieu}]{Sekse_2012}
Sekse, D.~H., Rouppe van~der Voort, L., \& De~Pontieu, B. 2013{\natexlab{a}},
  The Astrophysical Journal, 764, 164

\bibitem[{Sekse {et~al.}(2013{\natexlab{b}})Sekse, Rouppe van~der Voort,
  De~Pontieu, \& Scullion}]{Sekse_2013}
Sekse, D.~H., Rouppe van~der Voort, L., De~Pontieu, B., \& Scullion, E.
  2013{\natexlab{b}}, The Astrophysical Journal, 769, 44

\bibitem[{Shetye {et~al.}(2016)Shetye, Doyle, Scullion, Nelson, Kuridze,
  Henriques, Woeger, \& Ray}]{Shetye_2016}
Shetye, J., Doyle, J.~G., Scullion, E., {et~al.} 2016, Astronomy \&
  Astrophysics, 589, A3

\bibitem[{Shibata {et~al.}(2007)Shibata, Nakamura, Matsumoto, Otsuji, Okamoto,
  Nishizuka, Kawate, Watanabe, Nagata, UeNo, Kitai, Nozawa, Tsuneta, Suematsu,
  Ichimoto, Shimizu, Katsukawa, Tarbell, Berger, Lites, Shine, \&
  Title}]{Shibata_2007}
Shibata, K., Nakamura, T., Matsumoto, T., {et~al.} 2007, Science, 318,
  1591–1594

\bibitem[{{Sollum}(1999)}]{Sollum_1999}
{Sollum}, E. 1999, Master's thesis, University of Oslo, Institute of
  Theoretical Astrophysics

\bibitem[{Srivastava {et~al.}(2017)Srivastava, Shetye, Murawski, Doyle,
  Stangalini, Scullion, Ray, W{\'o}jcik, \& Dwivedi}]{Srivastava_2017}
Srivastava, A.~K., Shetye, J., Murawski, K., {et~al.} 2017, Scientific Reports,
  7, 43147

\bibitem[{{Sterling}(1998)}]{Sterling_1998}
{Sterling}, A. 1998, in ESA Special Publication, Vol. 421, Solar Jets and
  Coronal Plumes, ed. T.-D. {Guyenne}, 35

\bibitem[{{Uitenbroek}(2001)}]{Uitenbroek_2001}
{Uitenbroek}, H. 2001, \apj, 557, 389

\bibitem[{{V{\"o}gler} {et~al.}(2005){V{\"o}gler}, {Shelyag}, {Sch{\"u}ssler},
  {Cattaneo}, {Emonet}, \& {Linde}}]{Voegler_2005}
{V{\"o}gler}, A., {Shelyag}, S., {Sch{\"u}ssler}, M., {et~al.} 2005, \aap, 429,
  335

\bibitem[{{White} \& {Wilson}(1966)}]{White_1966}
{White}, O.~R. \& {Wilson}, P.~R. 1966, \apj, 146, 250

\end{thebibliography}
\end{document}